\documentclass[usenatbib]{mnras}
\usepackage{amssymb,amsmath}
\usepackage{graphicx}
\usepackage{xcolor,natbib}
\usepackage{multirow}
\usepackage[T1]{fontenc}
\usepackage{ae,aecompl}
\usepackage{newtxtext, newtxmath}
\usepackage{float}
\usepackage{subfigure}
\usepackage{longtable}
\usepackage{enumitem}
\usepackage{longtable,listings}
\usepackage[flushleft]{threeparttable}
\usepackage{parcolumns}
\usepackage{makecell}

\def\o3c{[O~{\sc iii}]$_c$}
\def\o3b{[O~{\sc iii}$_B$]}
\def\dif{\mathop{}\!\mathrm{d}}

\title[clues for hidden TDEs in AGN]{Interesting Clues to Detect Hidden Tidal Disruption Events in Active Galactic Nuclei}
\author[Zhang]
{Xue-Guang Zhang\thanks{Corresponding author Email:
            \href{mailto:xgzhang@gxu.edu.cn}{xgzhang@gxu.edu.cn}} \\
Guangxi Key Laboratory for Relativistic Astrophysics, School of Physical Science and Technology, 
GuangXi University, Nanning, 530004, P. R. China}

\date{}

\begin{document}
\pagerange{\pageref{firstpage}--\pageref{lastpage}} \pubyear{2023}
\maketitle
\label{firstpage}

\begin{abstract}
	In the manuscript, effects of Tidal Disruption Events (TDEs) are estimated on long-term AGN variability, to provide 
interesting clues to detect probable hidden TDEs in normal broad line AGN with apparent intrinsic variability which overwhelm 
the TDEs expected variability features, after considering the unique TDEs expected variability patterns. Based on theoretical 
TDEs expected variability plus AGN intrinsic variability randomly simulated by Continuous AutoRegressive process, long-term 
variability properties with and without TDEs contributions are well analyzed in AGN. Then, interesting effects of TDEs can be 
determined on long-term observed variability of AGN. First, more massive BHs, especially masses larger than $10^7{\rm M_\odot}$, 
can lead to more sensitive and positive dependence of $\tau_{TN}$ on $R_{TN}$, with $\tau_{TN}$ as variability timescale ratio 
of light curves with TDEs contributions to intrinsic light curves without TDEs contributions, and $R_{TN}$ as ratio of peak 
intensity of TDEs expected variability to the mean intensity of intrinsic AGN variability without TDEs contributions. Second, 
stronger TDEs contributions $R_{TN}$ can lead to $\tau_{TN}$ quite larger than 5. Third, for intrinsic AGN variability having 
longer variability timescales, TDEs contributions will lead $\tau_{TN}$ to be increased more slowly. The results actually provide 
an interesting forward-looking method to detect probable hidden TDEs in normal broad line AGN, due to quite different variability 
properties, especially different DRW/CAR process expected variability timescales, in different epochs, especially in normal broad 
line AGN with shorter intrinsic variability timescales and with BH masses larger than $10^7{\rm M_\odot}$. 
\end{abstract}


\section{Introduction}

	Variability is one of fundamental characteristics of active galactic nuclei (AGN) \citep{mr84, ww95, um97, ms16, db19, 
bv20, bs21}. Although there is uncertain physical origin of the AGN variability with different timescales, such as the proposed 
different models in \citet{tf00, ha02, fc05, lc08, pg13, ss18, pp22}, etc, there is a preferred mathematical process to describe 
the long-term AGN variability, damped random walk (DRW) process or Continuous AutoRegressive (CAR) process with two basic process 
parameters of intrinsic variability timescale $\tau$ and amplitude $\sigma$. \citet{kbs09} have firstly proposed the CAR process 
\citep{bd02} to described long-term AGN variability. And then, \citet{koz10, zu11, bj12, kb14, ss16, tm18, mv19, sr22} have 
provided improved methods to estimate the process parameters. 

	There are many reported studies on the AGN variability through the DRW process. \citet{mi10} have modeled the variability 
of about 9000 spectroscopically confirmed quasars covered in the SDSS Stripe82 region, and found correlations between the AGN 
parameters and the DRW process determined parameters. \citet{bj12} proposed an another fully probabilistic method for modeling 
AGN variability by the DRW process. \citet{ak13} have shown that the DRW process is preferred to model AGN variability, rather 
than several other stochastic and deterministic models, by fitted results of long-term variabilityof 6304 quasars. \citet{zu13} 
have checked that the DRW process provided an adequate description of AGN optical variability across all timescales. \citet{zh17} 
have checked long-term variability properties of AGN with double-peaked broad emission lines, and found the difference in 
intrinsic variability timescales between normal broad line AGN and the AGN with double-peaked broad emission lines. \citet{sl18} 
have modeled variability by DRW process and reported statistical analysis of the connection between AGN variability and physical 
properties of the central AGN activities, through the 2345 sources detected in both SDSS (Sloan Digital Sky Survey) and QUEST-La 
Silla. \citet{bs20} have modeled the month-long, 30 minute-cadence, high-precision TESS (Transiting Exoplanet Survey Satellite) 
light curve by the DRW process in the well-known archetypical dwarf AGN NGC 4395. More recently, \citet{si21} have modeled 
15years-long variability of 9248 quasars covered in SDSS stripe 82 region by combining the Pan-STARRS1 PS1 (Panoramic Survey 
Telescope and Rapid Response System 1 Survey) and SDSS light curves. \citet{zh21a} have modeled long-term variability of a composite 
galaxy to provide flues to support a true Type-2 AGN. Therefore, the long-term AGN variability have been well accepted to be 
mathematically modeled by the DRW/CAR process.

	Meanwhile, as discussed in \citet{me11, kv15, gw17, tg20, ss22}, intrinsic AGN variability deviations from the simple 
DRW description on short timescales, and also the estimated intrinsic variability timescale in the DRW process probably rises 
with increased baseline. However, in the manuscript, long-term variability not on short timescales but with the same length 
of time durations are mainly considered, therefore, neither variability on short timescales nor effects of different lengths 
of baseline are discussed in the manuscript.
	Besides the long-term intrinsic AGN variability well described by the CAR/DRW process, there is an unique kind of 
variability related to tidal disruption events (TDEs), which cannot intrinsically follow the CAR process expected variability 
properties, due to unique TDEs variability patterns. The well-known pioneer work on TDEs can be found in \citet{re88} and then 
followed in \citet{lu97, kh04, lk09, ce12, gr13, gm14, wy18, mg19, st19, pk20, lo21, zl21, zh22}, etc. The basic picture of a 
TDE is as follows. A star can be tidally disrupted by gravitational tidal force of a central massive black hole (BH), when it 
passing close to the central BH with a distance larger than event horizon of the BH but smaller than tidal disruption radius 
$R_{\rm T}=R_\star\times(\frac{M_{\rm BH}}{M_\star})^{1/3}$ with $R_\star$, $M_\star$ and $M_{\rm BH}$ as radius and mass of 
the being disrupted star and mass of central BH, respectively. The fallback materials can be accreted by the central massive 
BH, leading to time dependent TDEs variability roughly proportional to $t^{\sim~-5/3}$ at late times.

	More recent reviews on theoretical simulations and/or observational results on TDEs can be found in \citet{ko15, lf15, 
st19}. There are more than 100 TDE candidates reported in literature, see detail in \url{https://tde.space/}. Meanwhile, the 
well-known public sky survey projects have lead to more and more TDEs candidates detected, such as the TDEs candidates 
discovered through the known SDSS Stripe82 database in \citet{ve11}, through the known Catalina Sky Survey (CSS, \citet{dr09}) 
in \citet{dd11}, through the PanSTARRS (panoramic survey telescope and rapid response system) in \citet{gs12, ch14}, through 
the PTF (palomar transient factory) in \citet{bl17, vv19}, through the Optical Gravitational Lensing Experiment (OGLE) in 
\citet{wz17, gr19}, through the ASAS-SN (all-sky automated survey for supernovae) in \citet{ht14, ho16, hi21}, through the CNSS 
(Caltech-NRAO Stripe 82 Survey) in \citet{an20}, and through the ZTF (Zwicky Transient Facility) in \citet{vv19, lh20, sv21}, 
etc. More recently, two large samples of dozens of new TDE candidates can be found in \citet{vg21} from the First Half of ZTF 
(Zwicky Transient Facility) Survey observations along with Swift UV and X-ray follow-up observations and in \citet{sg21} from 
the SRG all-sky survey observations and then confirmed by optical follow-up observations. More recent review on observational 
properties of reported TDEs can be found in \citet{gs21}. However, among the reported TDEs candidates, there are few TDEs 
detected in normal broad line AGN with both apparent and strong intrinsic AGN variability.

        Among the reported TDE candidates, especially optical TDE candidates, strong broad Balmer and Helium emission lines 
are fundamental spectroscopic characteristics, however, the detected broad emission lines are not expected to be tightly related 
to normal broad line regions in normal broad line AGN, but to be related to disk-like structures from TDE debris. The known cases 
with broad emission lines in TDEs candidates can be found in SDSS J0159 as discussed in \citet{md15, zh21b}, ASASSN-14li as 
discussed in \citet{ho16}, PTF09djl as discussed in \citet{lz17}, PS18kh as discussed in \citet{ht19}, AT2018hyz as discussed 
in \citet{sn20, hf20}, etc., indicating the reported broad emission lines in the TDE candidates are not related to normal BLRs 
in normal broad line AGN, but are tightly related to TDE debris. Moreover, there are several TDE candidates, their UV-band 
spectra have been well checked, such as the PS18kh, ASASSN-15lh, ASASSN-14li, etc., there are no broad Mg~{\sc ii}$\lambda2800$\AA~ 
emission lines. And moreover, in the TDEs candidates with detected optical broad emission lines, there are no clues on DRW 
process expected variability, except the TDEs expected variability patterns. In other words, there are no confirmed evidence 
to support central TDEs in normal broad line AGN with apparent intrinsic AGN variability.

	Certainly, not similar as in quiescent galaxies, a moving star can be tidally disrupted by the central supermassive BH 
without a pre-existing accretion disk. However, there is also an existed accretion disk around the central supermassive BH in 
AGN, therefore, effects of the existed accretion disk should be considered on accreting fallback TDEs debris in normal broad line 
AGN. \citet{kb17} have discussed effects of a pre-existing accretion disc on TDEs expected variability, leading to still TDE 
expected variability patterns but with a probable cut-off. \citet{cp19, cp20} have modeled TDEs variability  in AGN with a 
pre-existing accretion disc, and discussed evolutions of the fallback bound debris being modified by collisions with the 
pre-existing disk, indicating the expected variability should be not totally similar as the TDEs expected variability patterns. 
However, there are so-far several TDEs candidates detected and reported in AGN. \citet{bn17} have reported a TDE candidate in a 
narrow line Seyfert 1 galaxy of which light curves can be roughly described by theoretical TDE model, and discussed that out-of-plane 
TDEs have quite weak interactions between the TDE debris and the pre-existing disk because the debris only intersect a small 
region of the disk. \citet{yx18} have shown the TDE expected variability pattern in the low-luminosity AGN NGC 7213. \citet{zd20} 
have reported a TDE candidate in AGN SDSS J0227 with probable broad Balmer emission lines, and shown the sudden rise followed 
by a smooth decline trend in long-term variability in SDSS J0227. \citet{zs22} have shown the TDE expected variability patterns 
in a narrow line Seyfert 1 galaxy. More recently, \citet{zh22b} have shown TDE expected long-term variability in the high 
redshift quasar SDSS J014124+010306, and \citet{zh22c} have discussed and shown TDE expected long-term variability of broad 
H$\alpha$ line luminosity in low luminosity broad line AGN NGC 1097. Therefore, totally similar TDE simulating variability can 
be expected in normal AGN with pre-existing accretion disks.

	Rare TDEs reported in normal AGN are mainly due to stronger intrinsic AGN variability than TDEs variability. However, 
there are enough probabilities and feasibilities to expect TDEs in normal AGN with supermassive BHs, even there are no detected 
TDEs expected variability features which are probably overwhelmed by strong intrinsic AGN variability in observed light curves. 
For intrinsic long-term AGN variability, the expected timescales are simply consistent with accretion disk orbital timescales 
or thermal timescales of about hundreds of days as the shown results in \citet{kbs09} for normal AGN (including 55 AGN from the 
MACHO survey, 37 Palomar Green quasars, and eight Seyfert galaxies from the AGN Watch project), in \citet{koz10} for about 2700 
OGLE quasars, in \citet{mi10} for about 9000 quasars covered in the SDSS Stripe82 region, and in \citet{rs18} for extreme 
variability quasars. Meanwhile, for variability from probable TDEs around supermassive BHs with masses around 
$10^{7-8}{\rm M_\odot}$ in AGN, the expected years-long timescales can be compared to the timescales of intrinsic long-term AGN 
variability. Therefore, it is interesting to check effects of TDEs on long-term AGN variability, which could provide interesting 
clues to expect probable hidden TDEs in normal broad line AGN with CAR process described intrinsic variability, through the 
long-term light curves from the public sky survey projects. Section 2 and Section 3 present our main hypotheses and main results. 
Section 4 gives the discussions and further applications. Section 5 gives our final conclusions. And in the manuscript, the 
cosmological parameters of $H_{0}=70{\rm km\cdot s}^{-1}{\rm Mpc}^{-1}$, $\Omega_{\Lambda}=0.7$ and $\Omega_{\rm m}=0.3$ have 
been adopted.

\section{Main Hypotheses} 

\subsection{Time Dependent Bolometric Luminosities from TDEs}

	In the manuscript, the well discussed theoretical TDEs model in \citet{gr13, gm14, mg19} is mainly considered and accepted, 
combining with the mass-radius relation in \citet{tp96} accepted for main-sequence stars. The time dependent bolometric 
luminosities from TDEs are simulated by the following four steps, similar as what we have done in \citet{zh22} to model X-ray 
variability of the relativistic TDE candidate Swift J2058.4+0516 and in \citet{zh22b, zh22c} to model optical variability in quasar 
SDSS J014124+010306 and in low luminosity broad line AGN NGC 1097.

	First, standard templates of viscous-delayed accretion rates in TDEs are created. Based on both the TDEFIT 
(\url{https://tde.space/tdefit/}) code \citep{gm14} and the MOSFIT (Modular Open Source Fitter for Transients, 
\url{https://mosfit.readthedocs.io}) code \citep{gn18} provided $dm/de$ ($m$ as debris mass and $e$ the specific binding energy), 
templates of fallback material rate $\dot{M}_{fbt}~=~dm/de~\times~de/dt$ can be created for standard cases with the central BH 
mass $M_{\rm BH}=10^6{\rm M_\odot}$ and the being disrupted main-sequence star of $M_{*}=1{\rm M_\odot}$ and with a grid of the 
listed impact parameters $\beta_{t}$ in \citet{gr13}. Considering the viscous delay as discussed in \citet{gr13, mg19} by a 
parameter of viscous timescale $T_{vis}$, templates of viscous-delayed accretion rates $\dot{M}_{at}$ can be determined by
\begin{equation}
\dot{M}_{at}~=~\frac{exp(-t/T_{vis})}{T_{vis}}\int_{0}^{t}exp(t'/T_{vis})\dot{M}_{fbt}dt'
\end{equation}
A grid of 31 evenly distributed $\log(T_{vis,~t}/{\rm years})$ range from -3 to 0 are applied to create templates $\dot{M}_{at}$ 
for each impact parameter $\beta_t$. Therefore, the created templates $\dot{M}_{at}$ include 736 (640) time-dependent 
viscous-delayed accretion rates for 31 different $T_{vis}$ of each 23 (20) impact parameters $\beta_t$ for the main-sequence 
star with polytropic index $\gamma$ of 4/3 (5/3).

	Second, for TDEs with model parameters of $\beta$ and $T_{vis}$ different from the list values in $\beta_{t}$ and in 
$T_{vis,~t}$, the corresponding viscous-delayed accretion rates $\dot{M}_{a}$ are created by the following two line interpretations. 
Assuming that $\beta_1$, $\beta_2$ in the $\beta_{t}$ are the two values nearer to the input $\beta$, and that $T_{vis1}$, 
$T_{vis2}$ in the $T_{vis,~t}$ are the two values nearer to the input $T_{vis}$, the first linear interpretation is applied to 
find the viscous-delayed accretion rates with input $T_{vis}$ but with $\beta~=~\beta_1$ and $\beta~=~\beta_2$ by
\begin{equation}
\begin{split}
&\dot{M}_{a}(T_{vis},~\beta_{1})~=~\dot{M}_{at}(T_{vis1},~\beta_1) + \\
&\ \ \ \ \ \frac{T_{vis}-T_{vis1}}{T_{vis2}-T_{vis1}}(\dot{M}_{at}(T_{vis2},~\beta_1)
	- \dot{M}_{at}(T_{vis1}, \beta_1))\\
&\dot{M}_{a}(T_{vis},~\beta_2)~=~\dot{M}_{at}(T_{vis1},~\beta_2) + \\
&\ \ \ \ \ \frac{T_{vis}-T_{vis1}}{T_{vis2}-T_{vis1}}(\dot{M}_{at}(T_{vis2},~\beta_2)
	- \dot{M}_{at}(T_{vis1},~\beta_2))
\end{split}
\end{equation}
The second linear interpretation is applied to find the viscous-delayed accretion rates with input $T_{vis}$ and with input $\beta$ by
\begin{equation}
\begin{split}
&\dot{M}_{a}(T_{vis},~\beta)~=~\dot{M}_{a}(T_{vis},~\beta_1) + \\
&\ \ \ \ \ \frac{\beta-\beta_1}{\beta_2-\beta_1}(\dot{M}_{a}(T_{vis},~\beta_2)
	- \dot{M}_{a}(T_{vis},~\beta_1))
\end{split}
\end{equation}

	Third, for TDEs with input parameters of $M_{\rm BH}$ and $M_{*}$ different from $10^6{\rm M_\odot}$ and $1{\rm M_\odot}$, 
the viscous-delayed accretion rates $\dot{M}$ and the corresponding time information $t$ in observer frame are created by the 
following scaling relations as shown in \citet{gm14, mg19},
\begin{equation}
\begin{split}
&\dot{M}~=~M_{\rm BH,~6}^{-0.5}~\times~M_{\star}^2~\times~
	R_{\star}^{-1.5}~\times~\dot{M}_{a}(T_{vis},~\beta) \\
&t~=~(1+z)\times M_{\rm BH}^{0.5}~\times~M_{\star}^{-1}\times
	R_{\star}^{1.5}~\times~t_{a}(T_{vis},~\beta)
\end{split}
\end{equation}
where $M_{\rm BH,~6}$, $M_{\star}$, $R_{\star}$ and $z$ represent central BH mass in unit of ${\rm 10^6M_\odot}$, stellar mass 
in unit of ${\rm M_\odot}$, mass-radius relation determined stellar radius in unit of ${\rm R_{\odot}}$, and redshift of host 
galaxy, respectively.

	Fourth, the time dependent bolometric luminosities $L_{\rm bol,~t,~TDE}$ from TDEs can be finally calculated by 
\begin{equation}
L_{\rm bol,~t,~TDE}~=~\eta~\times~\dot{M}(t)~c^2
\end{equation}
where $c$ and $\eta$ are the light speed and the energy transfer efficiency around central BH. The value $\eta$ will be further 
discussed in the following subsections. Therefore, for a TDE with given model parameters of central BH mass $M_{BH}$, stellar 
mass $M_\star$ and polytropic index $\gamma$ of the central being disrupted main-sequence star, the impact parameter $\beta$, 
the viscous timescale $T_{vis}$, redshift $z$ and energy transfer efficiency $\eta$, time dependent $L_{\rm bol,~t,~TDE}$ can 
be well simulated by the theoretical TDEs model.

	Based on the four steps, TDE expected time dependent bolometric luminosities can be simulated by accepted the only 
criterion that the TDE model parameters determined tidal radius larger than the event horizon of central BH.

	Before the end of the subsection, two points are noted. First, the circularizations in TDEs as discussed in \citet{kc94, 
br16, hs16, zo20, lo21} are not considered in the manuscript. The circularization emissions in TDEs have been probably detected 
in the TDE candidate ASASSN-15lh in \citet{lf16} and in TDE candidate AT 2019avd in \citet{cd22}, due to the two clear peaks (or 
two clear phases) detected in the NUV and/or optical band light curves. However, among the more than 100 reported TDEs candidates, 
there are rare TDEs candidates of which optical light curves have re-brightened peaks, indicating the ratio of TDEs with clear 
circularization emissions is very low. Therefore, we mainly consider the simple case that the fallback timescales of the 
circularizations are significantly smaller than the viscous timescales of the accretion processes, and the fallback materials 
will circularize into a disk as soon as possible. Second, the expected plateau phase in TDEs expected light curves with 
considerations of pre-existing accretion disk of AGN are not considered in the manuscript, because the plateau phase has small 
time duration and/or no plateau phases in some AGN (such as the results in \citet{yx18, zs22, zh22b, zh22c}, etc.) due to low 
surface density of pre-existing accretion disk of AGN.

\subsection{Time Dependent Bolometric Luminosities from the well-known AGN NGC5548}

	In the manuscript, the observed long-term light curve $L_{\rm c,~t,~N5548}$ of continuum luminosity at 5100\AA~ over 
13 years of the well-known broad line AGN NGC5548 ($z~=~0.01717$) in \citet{pe02} and in the AGNWATCH project 
(\url{https://www.asc.ohio-state.edu/astronomy/agnwatch/n5548/lcv/}) is collected as the AGN variability template. Then, the 
time dependent bolometric luminosity from NGC5548 $L_{\rm bol,~t,~N5548}~=~10~\times~L_{\rm c,~t,~N5548}$ is calculated by the 
bolometric corrections. The bolometric correction factor 10 is accepted, based on the statistical properties of spectral energy 
distributions of broad line AGN discussed in \citet{rg06, du20} and also on the more recent discussed results in \citet{nh20}. 

	Moreover, based on the well discussed results in \citet{pf04, bw10, pb14}, the central BH mass can be accepted as 
$M_{BH}\sim6.7\times10^{7}{\rm M_\odot}$\footnote{BH mass values varying from $2\times10^{7}{\rm M_\odot}$ to 
$8\times10^{7}{\rm M_\odot}$ in NGC5548 have few effects on our final results.} in the well-known reverberation mapped broad 
line AGN NGC5548 in the AGNWATCH project and in the LAMP (Lick AGN Monitoring Project) project 
(\url{https://www.physics.uci.edu/~barth/lamp.html}). And \citet{ld16} have reported similar BH mass of NGC5548 by the 
reverberation mapped results through Lijiang 2.4m telescope at Yunnan Observatory. More recently, \citet{wp20, hd21} have 
reported similar BH mass of NGC5548, through the space telescope and optical reverberation mapping project. Then, based on 
the well discussed results in \citet{dl11}, the energy transfer efficiency around the central BH in NGC5548 can be well 
estimated as 
\begin{equation}
\eta~=~0.089\times(\frac{M_{BH}}{\rm 10^8 M_\odot})^{0.52}~=~7.2\%
\end{equation}
which will be applied in Equation (5) above.

\begin{figure}
\centering\includegraphics[width = 8cm,height=5cm]{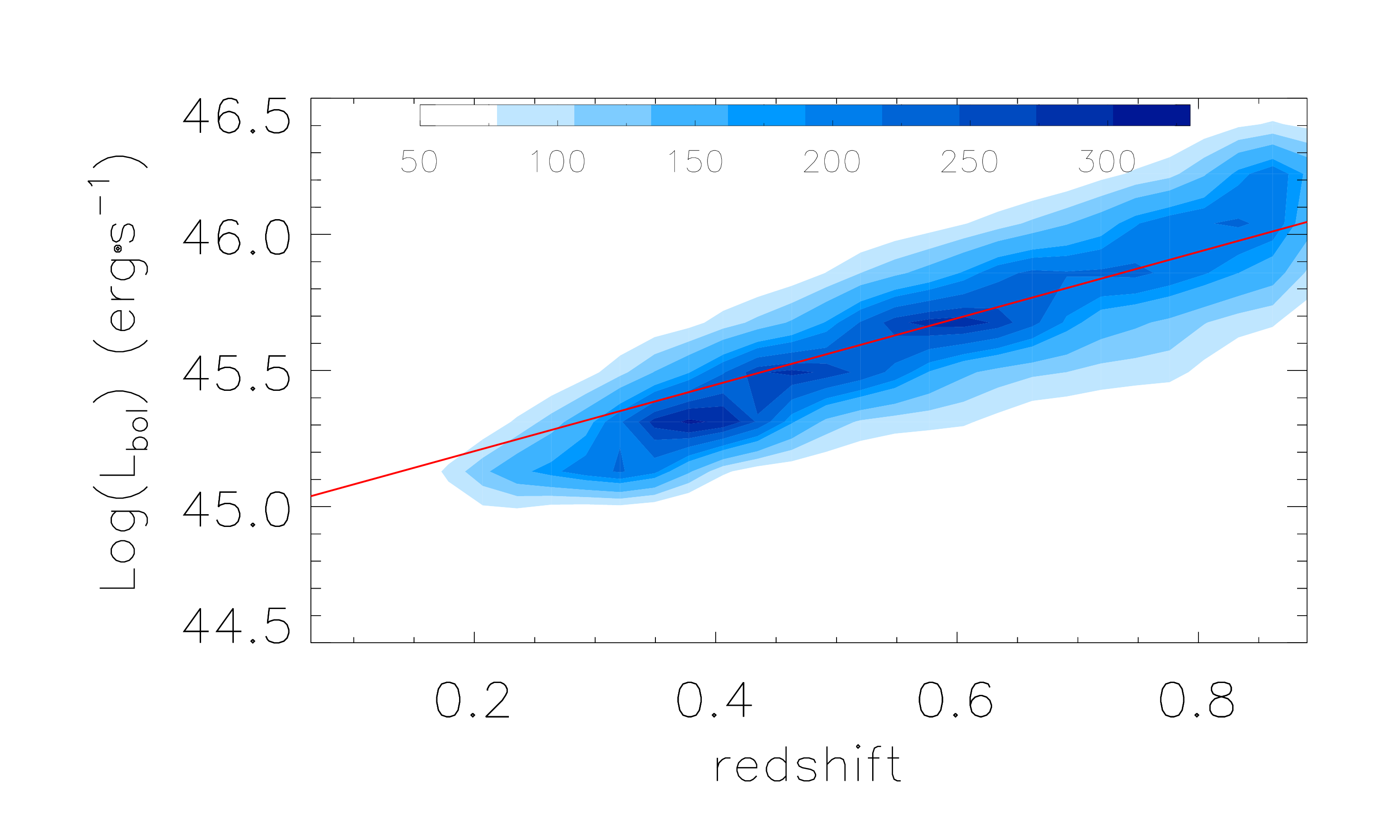}
\caption{Dependence of bolometric luminosity (10 times of the continuum luminosity at 5100\AA) on redshift for all 
the collected SDSS quasars with reliable measurements of continuum luminosity from \citet{sh11}. Solid red line shows the 
best description $L_{b0}~=~44.96~+~1.22~\times~z$.
}
\label{Bz}
\end{figure}

\begin{figure*}
\centering\includegraphics[width = 18cm,height=10cm]{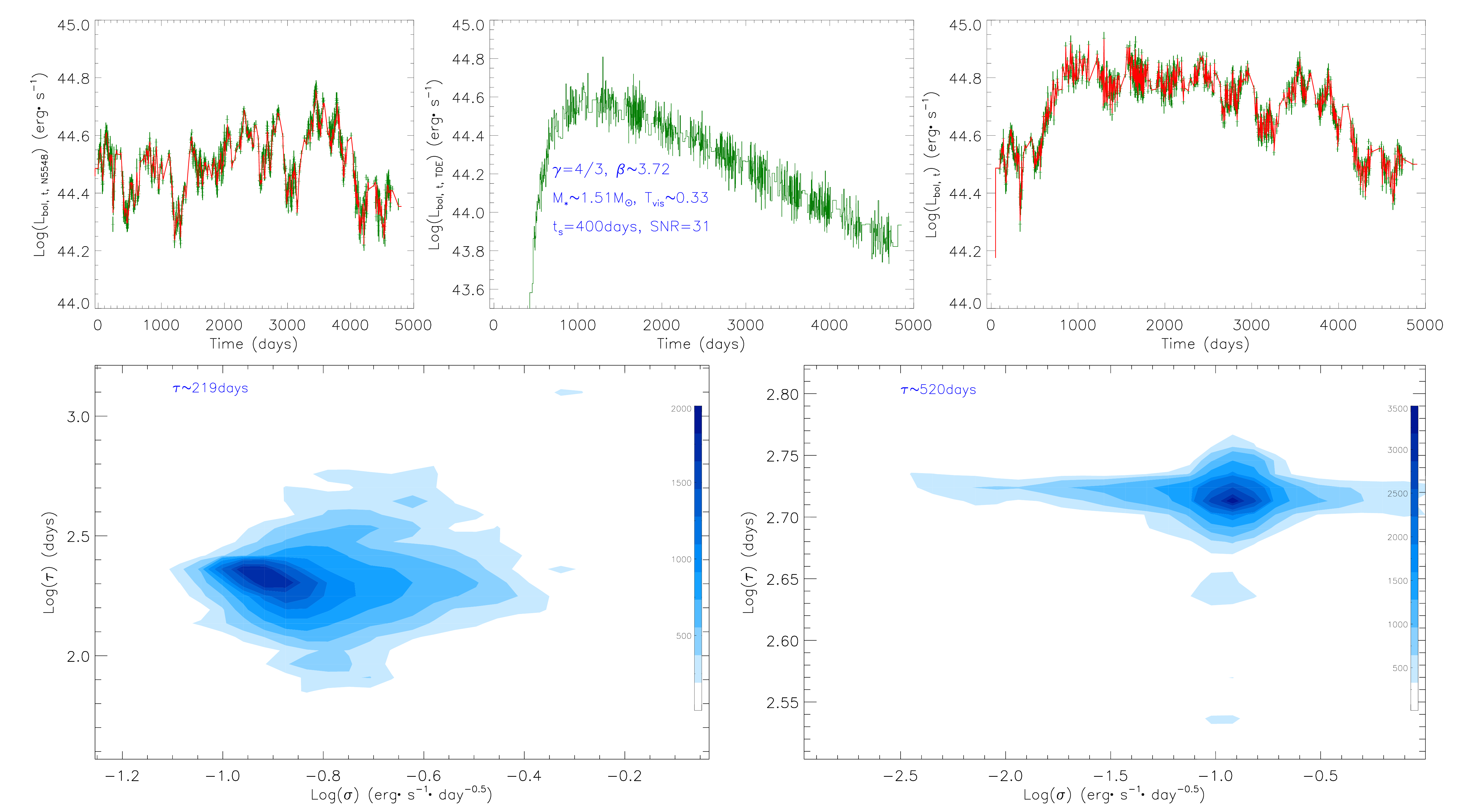}
\caption{Top left panel shows $L_{\rm bol,~t,~N5548}$ of NGC5548 (in dark green) and the kbs09 method determined best descriptions 
(solid red line). Bottom left panel shows the MCMC technique determined two-dimensional posterior distributions in contour of 
$\sigma$ and $\tau$ of $L_{\rm bol,~t,~N5548}$. Top middle panel shows an example of mock TDEs light curve $L_{\rm bol,~t,~TDE}$ 
with model parameters marked in the panel. And due to small SNR, the light curve $L_{\rm bol,~t,~TDE}$ is not smooth. Top right 
panel shows an example of mock light curve $L_{\rm bol,~t}$ (solid circles plus error bars in dark green) by $L_{\rm bol,~t,~N5548}$ 
shown in the top left panel plus the $L_{\rm bol,~t,~TDE}$ shown in the top middle panel, and the kbs09 method determined best 
descriptions (solid red line). Bottom right panel shows the MCMC technique determined two-dimensional posterior distributions 
in contour of $\sigma$ and $\tau$ of $L_{\rm bol,~t}$ shown in the top-right panel.}
\label{n5548}
\end{figure*}

\begin{figure*}
\centering\includegraphics[width = 18cm,height=6cm]{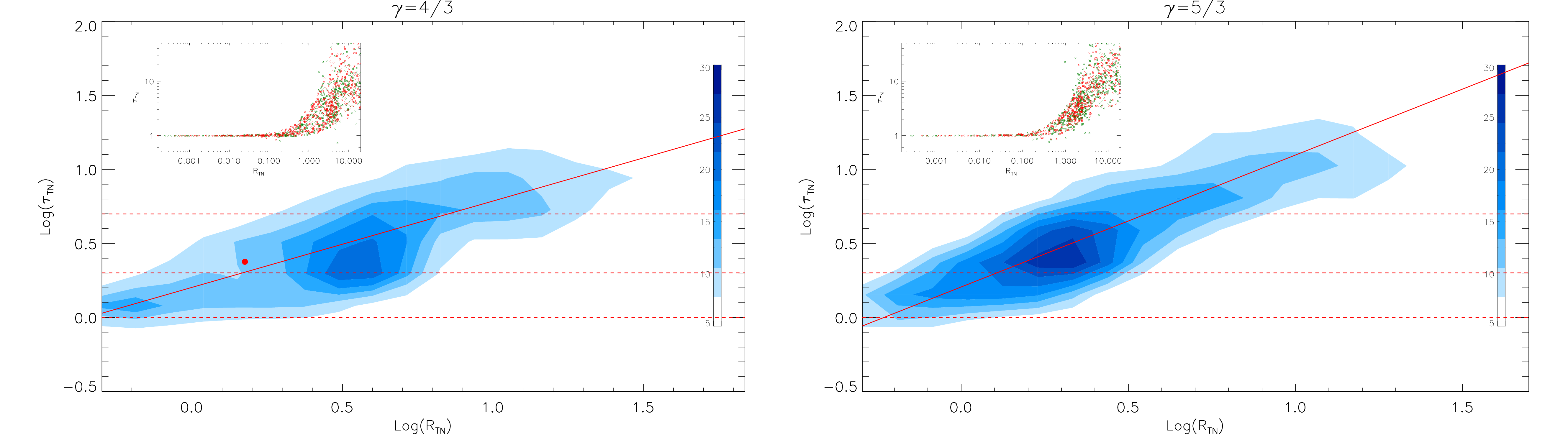}
\caption{Dependence of $\tau_{TN}$ on $R_{TN}$ and simple linear description in solid red line, based on the mock light curves 
$L_{bol,~t}$ created by $L_{\rm bol,~t,~N5548}$ plus contributions of TDEs with $\gamma~=~4/3$ (in the left panel), and with 
$\gamma~=~5/3$ (in the right panel). In left panel, solid red circle shows the results for the mock light curve $L_{bol,~t}$ 
shown in the top right panel of Fig.~\ref{n5548}. In each panel, top corner shows the results for all the 1200 mock light curves 
$L_{bol,~t}$, however the contour is plotted for the cases with $R_{TN}~>~0.5$. In each panel, from top to bottom, dashed red 
lines show $\tau_{TN}~=~5,~2,~1$, respectively. And in each top corner, symbols in red and in dark green show the cases with 
SNR larger than 55 and smaller than 55, respectively. Meanwhile, in each top corner, due to dense data points, the error bars 
with uncertainties about 20\% are not plotted.
}
\label{ns}
\end{figure*}

\subsection{Time Dependent Bolometric Luminosities with considerations of both AGN and TDE}

	There are three kinds of mock light curves $L_{\rm bol,~t}$ created by AGN intrinsic variability plus TDEs contributions, 
from simplicity to complexity. The first kind is to simply add mock light curve $L_{\rm bol,~t,~TDE}$ to the light curve 
$L_{\rm bol,~t,~N5548}$. The second kind is to add mock light curve $L_{\rm bol,~t,~TDE}$ to a randomly modified light curve 
$L_{\rm bol,~t,~AGN}$ which is created by $L_{\rm bol,~t,~N5548}$ plus a CAR process randomly created long-term variability. 
The third kind is created by CAR process randomly simulated long-term variability with different central physical properties.

	The first kind of $L_{\rm bol,~t}$ are simply created as follows. Mock light curves $L_{\rm bol,~t,~TDE}$ are created by 
randomly selected TDEs model parameters. The BH mass $M_{BH}$ and $\eta$ is fixed to $6.7\times10^{7}{\rm M_\odot}$ and $0.072$ 
(the values of NGC5548). The stellar mass $M_\star$ is randomly selected from $-2~<~\log(M_\star/{\rm M_\odot})~<~1$. The 
polytropic index $\gamma$ is selected to be 4/3 or 5/3. The impact parameter is randomly selected from the minimum $\beta_{t}$ 
to the maximum $\beta_{t}$. The viscous timescale $T_{vis}$ is randomly selected from the minimum $T_{vis,~t}$ to the maximum 
$T_{vis,~t}$. Here, there is a criterion that the expected tidal radius
\begin{equation}
\frac{R_{\rm TDE}}{R_{\rm s}}~=~5.06(M_\star)^{-1/3}
	(\frac{M_{BH,~6}}{10})^{-2/3}~R_\star~>~1
\end{equation}
larger than event horizon of central BH ($R_{\rm s}~=~2GM_{\rm BH}/c^2$). Then, with $\gamma~=~4/3$ ($\gamma~=~5/3$), 1200 
(1200) mock light curves $L_{\rm bol,~t,~TDE}$ are randomly created. Considering TDEs with different starting times $t_s$ 
randomly from 0 to 3000days, the mock $L_{\rm bol,~t}$ are created by 
\begin{equation}
L_{\rm bol,~t}~=~L_{\rm bol,~t+t_s,~TDE}~+~L_{\rm bol,~t,~N5548}
\end{equation} 
Then, different white noises defined by signal-to-noise ratio (SNR) randomly from 30 to 80 are added to the mock light curves 
$L_{\rm bol,~t}$. And the observational uncertainties of $L_{\rm bol,~t,~N5548}$ are accepted as the uncertainties of 
$L_{\rm bol,~t}$.

	Before proceeding further, simple discussions are given to describe why values of SNRs for white noises are randomly 
selected from 30 to 80. As the collected information of the long-term light curve of NGC5548, the mean ratio of continuum 
emissions to uncertainties of continuum emissions is about 32. Meanwhile, to our knowledge, among our collected low-redshift 
($z<0.35$) SDSS (Sloan Digital Sky Survey) quasars, such as the sample discussed in \citet{zh23}, the highest signal-to-noise 
ratio of SDSS spectra is about 74. Therefore, when adding white noises to the created mock light curves in the manuscript, 
corresponding SNRs are randomly selected from 30 to 80. Meanwhile, accepted SNRs from 30 to 80, corresponding photometric 
magnitude uncertainty can be simply estimated to be from 0.036mag to 0.013mag, which are similar as the magnitude uncertainties 
of light curves of quasars provided by SDSS Stripe82 database \citep{mi10}.

	The second kind of $L_{\rm bol,~t}$ are created as follows. The mock light curves $L_{\rm bol,~t+t_s,~TDE}$ are 
similarly created, but the AGN variability template $L_{\rm bol,~t,~AGN}$ is created by 
\begin{equation}
L_{\rm bol,~t,~AGN}~=~L_{\rm bol,~t,~N5548}~+~L(CAR)
\end{equation} 
where $L(CAR)$ is a randomly created light curve with mean of zero. And the $L(CAR)$ (with expected variance around 0.012) is 
randomly created through the CAR process described in \citet{kbs09}, 
\begin{equation}
\dif L(CAR)~=~\frac{-1}{\tau}L(CAR)\dif t~+~\sigma\sqrt{\dif t}\epsilon(t)
\end{equation}
where $\epsilon(t)$ a white noise process with zero mean and variance equal to 1. Here, the parameter $\tau$ is randomly selected 
from 100days to 1000days, as the shown results in \citet{mi10} for normal quasars. Then, the mock $L_{\rm bol,~t}$ are 
similarly created by
\begin{equation}
L_{\rm bol,~t}~=~L_{\rm bol,~t+t_s,~TDE}~+~L_{\rm bol,~t,~AGN}
\end{equation}
And different white noises defined by SNRs randomly from 30 to 80 are added to the mock light curves $L_{\rm bol,~t}$. Here, the 
light curve $L_{\rm bol,~t,~AGN}$ has different intrinsic variability timescale and amplitude from those of$L_{\rm bol,~t,~N5548}$, 
which will provide further considerations of effects of TDEs on long-term variability of AGN. And the observational uncertainties 
of $L_{\rm bol,~t,~N5548}$ are accepted as the uncertainties of $L_{\rm bol,~t}$.

	The third kind of $L_{\rm bol,~t}$ is mainly created as follows after considering different parameters of BH mass, redshift, 
energy transfer efficiency, etc. The AGN variability template $L_{\rm bol,~t,~CAR}$ is created by the CAR process determined 
$L(CAR)$ plus an expected bolometric luminosity $L_{b0}$ ($\log(L_{bol}/{\rm erg/s})$) depending on redshift,
\begin{equation}
\begin{split}
&L_{\rm bol,~t,~CAR}~=~L_{b0}~+~L(CAR) \\
&\dif L(CAR)~=~\frac{-1}{\tau_0}L(CAR)\dif t~+~\sigma\sqrt{\dif t}\epsilon(t) \\
&L_{b0}~=~44.96~+~1.22~\times~z
\end{split}
\end{equation}
where $\tau_0$ is selected to be 200days or 600days (a common value and a large value of intrinsic variability timescale in quasars, 
see results in \citet{mi10, kbs09, koz10, rs18}), and $\frac{\sigma^2\tau}{2}$ is selected to be around 0.012 (leading to similar 
variance as those in NGC5548). The selected parameters of $\tau$ and $\sigma$ lead the $L(CAR)$ with mean of zero and variance 
similar as the $L_{\rm bol,~t,~N5548}$. The dependence of bolometric luminosity on redshift $L_{b0}~\propto~1.22~\times~z$, 
shown in Fig,~\ref{Bz}, is well determined from all the 23093 SDSS quasars in \citet{sh11} with measured continuum luminosity at 
5100\AA. There is a strong positive correlation between redshift and bolometric luminosity calculated by 10 times of the continuum 
luminosity at 5100\AA, with Spearman rank correlation coefficient 0.66 ($P_{null}~<~10^{-15}$) and with RMS scatter about 0.29. 
Here, 6 different values of 0.05,~0.1,~0.2,~0.3,~0.5,~1 are accepted as input redshift, applied to determine $L_{0}$. Meanwhile, 
based on the three different BH masses $M_{BH}~=~10^6,~10^7,~5\times10^7~{\rm M_\odot}$, three different energy transfer efficiency 
$\eta~=~0.06,~0.15,~0.3$ and the six redshift, the $L_{\rm bol,~t+t_s,~TDE}(M_{BH},~\eta,~z)$ can be randomly created. Then, the 
mock light curves $L_{\rm bol,~t}$ are similarly created by
\begin{equation}
L_{\rm bol,~t}~=~L_{\rm bol,~t+t_s,~TDE}(M_{BH},~\eta,~z)~+~L_{\rm bol,~t,~CAR}
\end{equation}
And different white noises defined by SNRs randomly from 30 to 80 are added to the mock light curves $L_{\rm bol,~t}$. For each 
series [$\gamma$,~$M_{BH}$,~$\eta$,~$z$,~$\tau$], 1200 mock light curves are created with contributions of TDEs. Finally, there 
are $2\times3\times3\times6\times2\times1200~=~259200$ mock light curves created after considering TDEs contributions to intrinsic 
AGN variability. And 10\% are accepted as the uncertainties of $L_{\rm bol,~t}$.

	Actually, besides the linear dependence of bolometric luminosity on redshift, dependence of BH mass on redshift is also 
checked through the reported parameters of the quasars in \citet{sh11}. However, the Spearman Rank correlation coefficient for 
the dependence is only 0.29, quite weaker than the dependence of bolometric luminosity on redshift. Therefore, rather than 
dependence of BH mass on redshift, the linear dependence of bolometric luminosity on redshift is accepted in the manuscript. The 
application of the linear dependence of bolometric luminosity on redshift can reduce one free model parameter to create the third 
kind of mock light curves. Moreover, as shown in \citet{mi10, kbs09}, there is a dependence of process parameter $\tau$ on BH 
mass. However, the dependence is very loose, with Spearman Rank correlation coefficient about 0.23. Therefore, in the manuscript, 
the loose dependence of process parameter $\tau$ on BH mass is not accepted. And accepted the BH mass and $\tau$ and redshift 
are independent parameters, much wider parameter space can be occupied to create the mock light curves, and more efficient 
conclusions can be obtained.

\begin{table*}
\caption{Model Parameters applied to create the three kinds of mock light curves}
\begin{tabular}{c|ccccccc|cc}
\hline\hline
	& \multicolumn{7}{c|}{parameters applied in TDE model with $\gamma=4/3, 5/3$} & \multicolumn{2}{c}{parameters for $L(CAR)$} \\
	& $M_{BH}$ & $\log(M_\star)$ & $\eta$  & $z$ & $\beta$ & $\log(T_{vis})$ & $t_s$ &  $\tau$  & $\frac{\sigma^2\tau}{2}$\\ 
\hline
	1st  &  $6.7\times10^7$  & $\in$[-2, 1] & 0.072  & 0.01717   &  $\in$[$\beta_l$,$\beta_m$] & 
	$\in$[-3,0]  & $\in$[0, 3000]  &  \dots  &  \dots  \\
\hline
	2nd  &  $6.7\times10^7$  & $\in$[-2, 1] & 0.072  & 0.01717   & $\in$[$\beta_l$,$\beta_m$] & 
	$\in$[-3,0]  & $\in$[0, 3000] & $\in$[100, 1000] & $\in$[0.003,0.048] \\
\hline
	3rd  &  \makecell[c]{$\subset$[$10^6$, $10^7$,\\$5\times10^7$]} & $\in$[-2, 1] & \makecell[c]{$\subset$[0.06, 0.15,\\0.30]} & 
	     \makecell[c]{$\subset$[0.05, 0.1, 0.2, \\0.3, 0.5, 1.0]} & $\in$[$\beta_l$,$\beta_m$] & 
	$\in$[-3,0] & $\in$[0, 3000] & $\subset$[200, 600] & $\in$[0.003,0.048] \\
\hline\hline
\end{tabular} \\
\begin{itemize}
\item[1: ] The first column shows which kind of mock light curves, '1st' means the first kind of mock light curve 
$L_{\rm bol,~t}~=~L_{\rm bol,~t+t_s,~TDE}~+~L_{\rm bol,~t,~N5548}$, '2nd' means the second kind of mock light curve 
$L_{\rm bol,~t}~=~L_{\rm bol,~t+t_s,~TDE}~+~L_{\rm bol,~t,~N5548}~+~L(CAR)$ (with $L(CAR)$ as CAR process created variability), 
'3rd' means the third kind of mock light curve $L_{\rm bol,~t}~=~L_{\rm bol,~t+t_s,~TDE}(M_{BH},~\eta,~z)~+~L_{\rm bol,~t,~CAR}$.
\item[2: ] The second column, the third column, the fourth column, the fifth column, the sixth column, the seventh column and the 
eighth column show the parameters of BH mass in units of ${\rm M_\odot}$, logarithmic stellar mass in units of ${\rm M_\odot}$ 
energy transfer efficiency $\eta$, redshift $z$, $\beta$, logarithmic $T_{vis}$ in units of years and shifted time in units 
of days, applied in theoretical TDE model. 
\item[3: ] The last two panels show the CAR process parameters of $\tau$ in units of days and $\frac{\sigma^2\tau}{2}$ (expected 
variance of the CAR created light curve) applied to created light curves $L(CAR)$. 
\item[4: ] In each cell for the parameters, if there is only one value, meaning that the parameter is fixed to the listed value. 
\item[5: ] In each cell for the parameters, if the mathematical symbol $\in$ is used, meaning that the parameter is randomly selected 
from the minimum value to the maximum value listed in the square brackets following the mathematical symbol $\in$. 
\item[6: ] In each cell for the parameters, if the mathematical symbol $\subset$ is used, meaning that value of the parameter is chosen 
from the values listed in the square brackets following the mathematical symbol $\subset$.
\item[7: ] In the last column, the parameter $\frac{\sigma^2\tau}{2}$ shows the expected variance of the CAR process created light curve. 
	Based on the variance 0.012 of the light curve of NGC5548, the $\frac{\sigma^2\tau}{2}$ is accepted to be larger than 
	$0.25\times0.012$ and smaller than $4\times0.012$ for the created $L(CAR)$ in the second kind and the third kind of 
	mock light curves.
\item[8: ] In the fifth column, the $\beta$ is randomly selected from 0.6 to 4 if $\gamma=4/3$, and randomly selected from 0.5 to 2.5 if 
	$\gamma=5/3$.
\end{itemize}
\end{table*}

	Before the end of the section, three points are noted. First and foremost, in order to clearly show properties of model 
parameters applied to create TDEs contributions and to create $L(CAR)$, Table~1 shows the accepted values and/or accepted ranges 
of the applied model parameters. Besides, the main objective of the manuscript is to determine effects of TDEs contributions on 
observed long-term AGN variability from simplicity to complexity. Therefore, when the first kind and the second kind of mock light 
curves are created, the oversimplified procedure is firstly applied with the fixed BH mass (the BH mass of NGC5548), the fixed 
energy transfer efficiency (determined by the BH mass of NGC5548) and the fixed redshift (the redshift of NGC5548). Then, effects of 
randomly selected values of model parameters are considered through the third kind of mock light curves. Last but not the least, 
for the three kinds of mock light curves $L_{\rm bol,~t}$, the corresponding maximum BH mass is $6.7\times10^{7}{\rm M_\odot}$ (the 
BH mass of NGC5548), which is a large (near to the Hills mass limit) but reasonable value, see the maximum BH mass about 
$66\times10^{6}{\rm M_\odot}$ determined by the MOSFIT in TDEs candidates in \citet{mg19}. Meanwhile, when the third kind of 
mock light curves are created, the Equation (6) is not applied to determined energy transfer efficiency, after considering the 
listed values of $\eta$ in \citet{mg19} that high $\eta$ could be expected around central BH with masses around $10^6{\rm M_\odot}$. 
And also as the shown results in \citet{mg19}, the collected $\eta$ values from 0.06 to 0.3 are also reasonable to create time 
dependent TDEs expected bolometric luminosities for the third kind of mock light curves.

\section{Main Results} 

\subsection{Results based on the Long-Term Variabilities of $L_{\rm bol,~t,~N5548}$}

	As the discussed results in \citet{kbs09} (see their Fig.~4), the long-term variability $L_{\rm bol,~t,~N5548}$ of 
NGC5548 has intrinsic variability timescale about 214days. The same method as shown in Equation (7)-(12) in \citet{kbs09} (the 
kbs09 method) is applied to analyze variability of $L_{\rm bol,~t,~N5548}$, in order to ensure the applied kbs09 method in the 
manuscript is reliable. Here, rather than the public JAVELIN (Just Another Vehicle for Estimating Lags In Nuclei) code in 
\citet{zu11, zu13}, the kbs09 method is applied in the manuscript, due to the following main reason. For each mock light curve 
with about 1500 data points (time duration longer than 10 years), the kbs09 method running in Surface Studio2 can 
give the final best-fitting results in ten minutes through the Levenberg-Marquardt least-squares minimization technique (the 
known MPFIT package, \citealt{mpf09}), however, the JAVELIN code will give the final results in more than one hour.

	The $L_{\rm bol,~t,~N5548}$ is shown in top left panel of Fig.~\ref{n5548}, with the kbs09 method determined best 
descriptions through the Maximum Likelihood method combining with the Markov Chain Monte Carlo (MCMC) technique \citep{fh13}, 
with the kbs09 method determined process parameters through the MPFIT package accepted as starting values of the 
process parameters in the MCMC technique. The determined posterior distributions of the parameters of $\tau$ and $\sigma$ are 
shown in the bottom left panel of Fig.~\ref{n5548}, with accepted $\log(\tau/days)~\sim~2.34_{-0.076}^{+0.107}$ 
($\tau~\sim~219_{-36}^{+60}{\rm days}$) which is well consistent with the reported 214days in \citet{kbs09}. Therefore, the 
applied kbs09 method is reliable enough.

	Through the kbs09 method applied through the Levenberg-Marquardt least-squares minimization technique, 
variability properties, especially the CAR process parameters of $\sigma$ and $\tau$, can be well determined for the total 
2400 mock light curves $L_{\rm bol,~t}$ created by $L_{\rm bol,~t,~N5548}$ plus $L_{\rm bol,~t,~TDE}$. Top middle panel and 
top right panel of Fig.~\ref{n5548} show an example of $L_{\rm bol,~t,~TDE}$ and an example of $L_{\rm bol,~t}$. For the shown 
example in top right panel of Fig.~\ref{n5548} without clear TDEs expected variability features, the determined variability 
timescale is about 520days, as the shown posterior distributions in bottom right panel of Fig.~\ref{n5548} determined by 
MCMC technique applied in the kbs09 method, significantly longer than the intrinsic 219days of NGC5548, indicating TDEs 
contributions can lead to larger variability timescales.

	In order to show clearer effects of TDEs contributions, two parameters $R_{TN}$ and $\tau_{TN}$ are defined, $R_{TN}$ 
as ratio of the peak intensity of $L_{\rm bol,~t,~TDE}$ to the mean intensity of $L_{\rm bol,~t,~N5548}$, and $\tau_{TN}$ as 
ratio of the variability timescale of $L_{\rm bol,~t}$ to the intrinsic variability timescale 219days of $L_{\rm bol,~t,~N5548}$. 
Then, Fig.~\ref{ns} shows the dependence of $\tau_{TN}$ on $R_{TN}$ of the 2400 mock light curves $L_{\rm bol,~t}$, 1200 light 
curves based on the $L_{\rm bol,~t,~TDE}$ created with $\gamma~=~4/3$ and 1200 light curves based on the $L_{\rm bol,~t,~TDE}$ 
created with $\gamma~=~5/3$. For $R_{TN}~>~0.5$ (stronger TDEs contributions), there are positive correlations between 
$\tau_{TN}$ on $R_{TN}$, with the Spearman rank correlation coefficient is about 0.71 (0.79) with $P_{null}~<~10^{-15}$ for the 
cases with $\gamma~=~4/3$ ($\gamma~=~5/3$). Here, the critical value $R_{TN}~>~0.5$ is simply determined that the variance of 
$\tau_{TN}$ of the data points with $R_{TN}~>~0.5$ is at least 2000 times larger than the variance of $\tau_{TN}$ of the data 
points with $R_{TN}~<~0.5$. Actually, small different critical values from 0.5 have few effects on the discussed results. After 
considering the uncertainties in both coordinates, the positive dependence with $R_{TN}~>~0.5$ can be simply described by
\begin{equation}
\begin{split}
\log(\tau_{TN})(\gamma~=~4/3)~=~0.20~+~0.58\log(R_{TN}) \\
\log(\tau_{TN})(\gamma~=~5/3)~=~0.21~+~0.89\log(R_{TN})
\end{split}
\end{equation}
through the FITEXY code (\url{https://idlastro.gsfc.nasa.gov/ftp/pro/math/fitexy.pro} written by Frank Varosi) as discussed in 
\citet{tr02}. It is clear that longer variability timescales can be confirmed with larger TDEs contributions. And SNRs have few 
effects on the results, based on the shown results in each top corner in each panel of Fig.~\ref{ns}.

\begin{figure*}
\centering\includegraphics[width = 18cm,height=4cm]{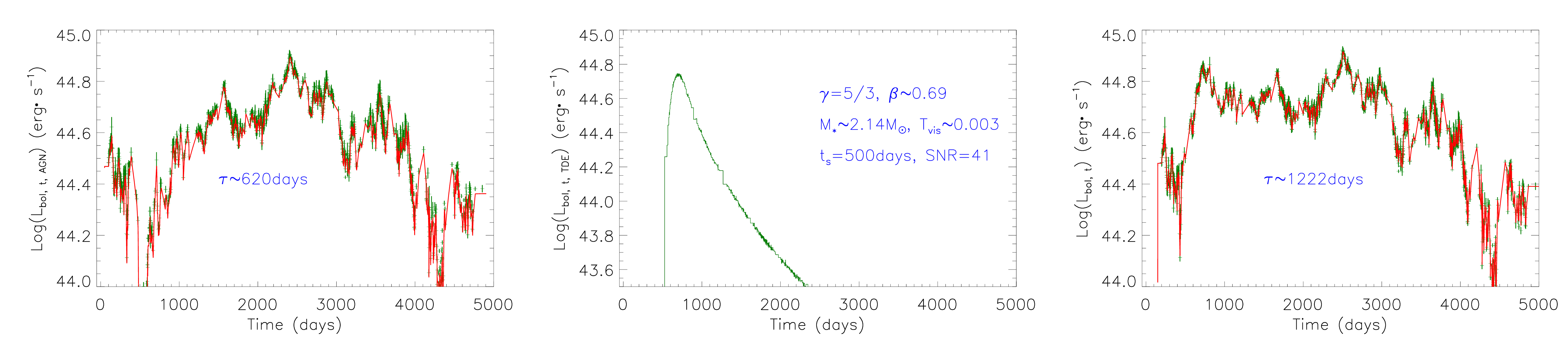}
\centering\includegraphics[width = 18cm,height=6cm]{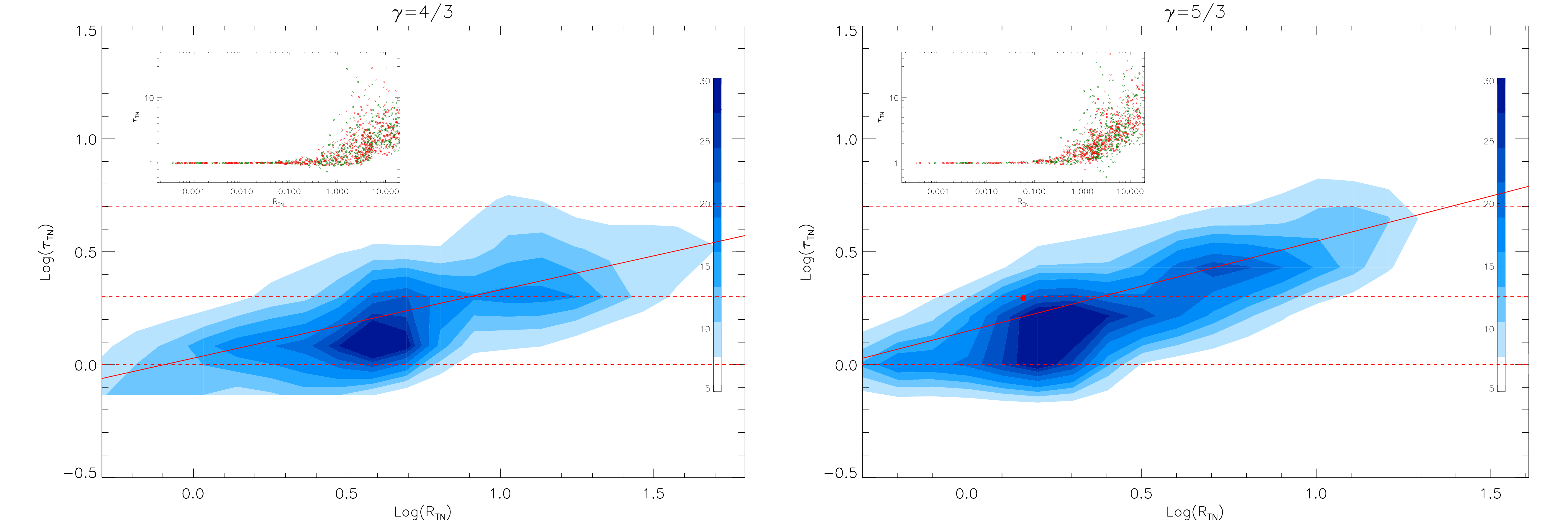}
\caption{Top panels show the results similar as those shown in top panels of Fig.~\ref{n5548}, but based on the light curve 
$L_{\rm bol,~t,~AGN}$ shown in the top left panel. Bottom panels show the results similar as those in Fig.~\ref{ns}, but based 
on the light curve $L_{\rm bol,~t,~AGN}$ with intrinsic variability timescale about 620days. In the bottom right panel, the 
solid red circle shows the results for the mock light curve $L_{bol,~t}$ shown in the top right panel. In top corners of bottom 
panels, due to large number of dense data points, the error bars with uncertainties about 20\% are not plotted.}
\label{agn}
\end{figure*}

	Before end of the subsection, scatters of $\tau_{TN}$ for given $R_{TN}$ can be simply discusses as follows. For smaller 
values of $R_{TN}$, TDEs contributions are very tiny, leading to few effects of TDEs contributions on determined $\tau$ in mock 
light curves, indicating tiny scatters of $\tau_{TN}$. However, for larger values of $R_{TN}$ leading to apparent TDEs contributions, 
different values of stellar mass and $\beta$ and large value $T_{vis}$ can lead to quite different time durations of TDE expected 
light curves with the same peak intensity. The different time durations can lead to different ratios of $\tau_{TN}$. Unless the 
model parameters applied in TDE model are fixed for a given $R_{TN}$, the scatters of $R_{TN}$ can be well expected, and provide 
robust clues in the manuscript to detect hidden TDEs in broad line AGN with apparent intrinsic variability. Similar scatters of 
$\tau_{TN}$ can also be expected in the following subsections.

\subsection{Results based on the Long-Term Variabilities of $L_{\rm bol,~t,~AGN}$}

	Similar as the results on $L_{\rm bol,~t,~N5548}$, top panels of Fig.~\ref{agn} show an example of mock light curve 
$L_{\rm bol,~t,~TDE}$ (in middle panel) and an example of mock light curve $L_{bol,~t}$ (in right panel) created by 
$L_{\rm bol,~t,~AGN}$ shown in the left panel plus the $L_{\rm bol,~t,~TDE}$ shown in the middle panel. And the kbs09 method 
is applied to determine the intrinsic variability timescale of $L_{\rm bol,~t,~AGN}$ as $\tau\sim620{\rm days}$  
through the Levenberg-Marquardt least-squares minimization technique. Bottom panels of Fig.~\ref{agn} shows the dependence 
of $\tau_{TN}$ on $R_{TN}$ of the 2400 mock light curves $L_{\rm bol,~t}$ based on the $L_{\rm bol,~t,~AGN}$. For $R_{TN}~>~0.5$, 
the Spearman rank correlation coefficient is about 0.63 (0.68) with $P_{null}~<~10^{-15}$ for the cases with 
$\gamma~=~4/3$ ($\gamma~=~5/3$). The positive dependence with $R_{TN}~>~0.5$ can be simply described by
\begin{equation}
\begin{split}
\log(\tau_{TN})(\gamma~=~4/3)~=~0.03~+~0.30\log(R_{TN}) \\
\log(\tau_{TN})(\gamma~=~5/3)~=~0.15~+~0.40\log(R_{TN})
\end{split}
\end{equation}
through the same FITEXY code.

	Similar results can be found that longer variability timescales can be confirmed with larger TDEs contributions, and 
SNRs have few effects on the results. However, the intrinsic AGN variability have longer variability timescales, the $\tau_{TN}$ 
will increase more slowly, based on the smaller slopes shown in the equations above.

\begin{figure*}
\centering\includegraphics[width = 18cm,height=8cm]{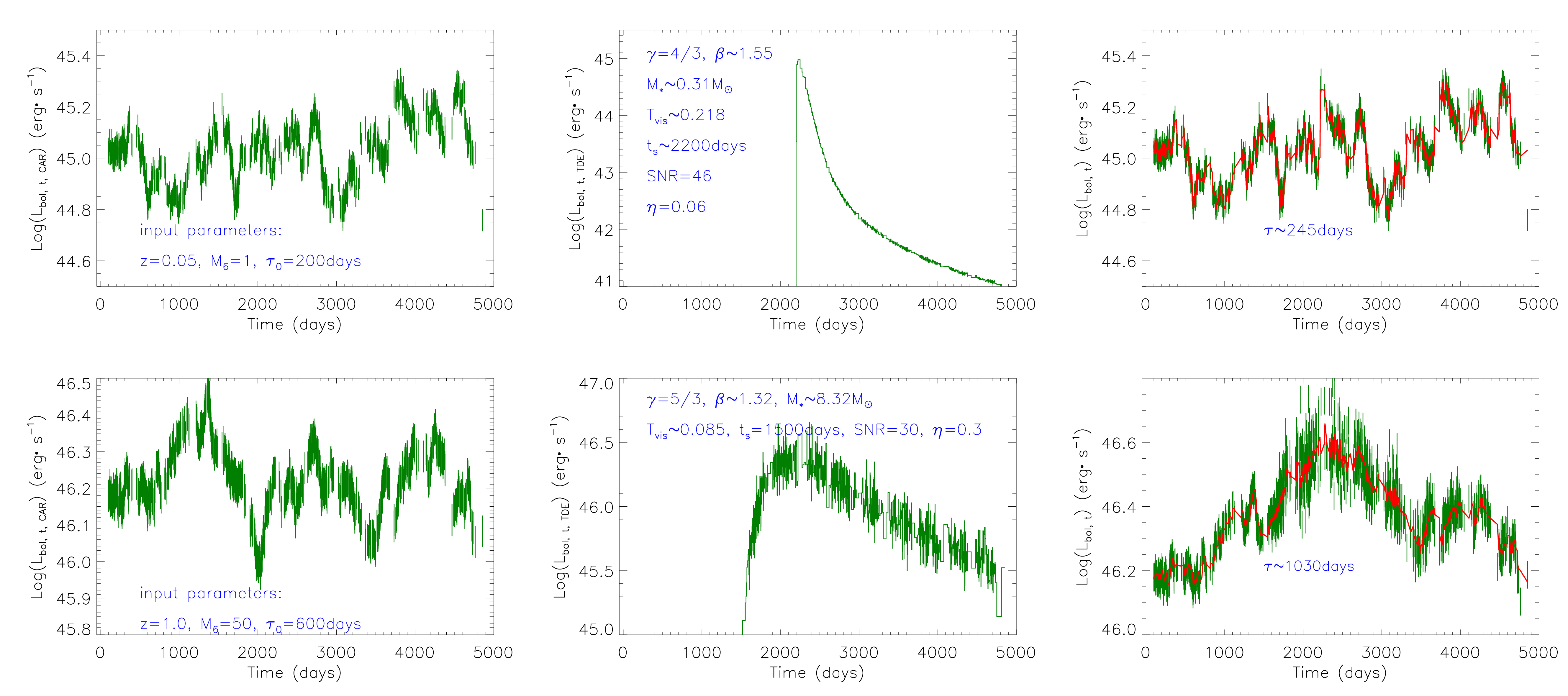}
\caption{Two examples on the mock light curves $L_{\rm bol,~t,~CAR}$ shown as dots plus error bars in dark green in left panels, 
the mock light curves $L_{\rm bol,~t,~TDE}(M_{BH},~\eta,~z)$ shown in the middle panels and the mock light curves $L_{\rm bol,~t}$ 
shown as dots plus error bars in dark green in the right panels. In each left panel, the input model parameters of BH mass $M_6$ 
(in unit of $10^6{\rm M_\odot}$), redshift, $\tau_0$ are listed in blue characters. In each middle panel, the input TDEs model 
parameters of $\gamma$, $\beta$, stellar mass $M_\star$, $T_{vis}$, $t_s$, $SNR$ and $\eta$ are listed in blue characters. In 
each right panel, solid red line shows the kbs09 method determined best descriptions to the $L_{\rm bol,~t}$, and the corresponding 
determined timescale $\tau$ is listed in blue characters.}
\label{car}
\end{figure*}

\begin{figure*}
\centering\includegraphics[width = 18cm,height=20cm]{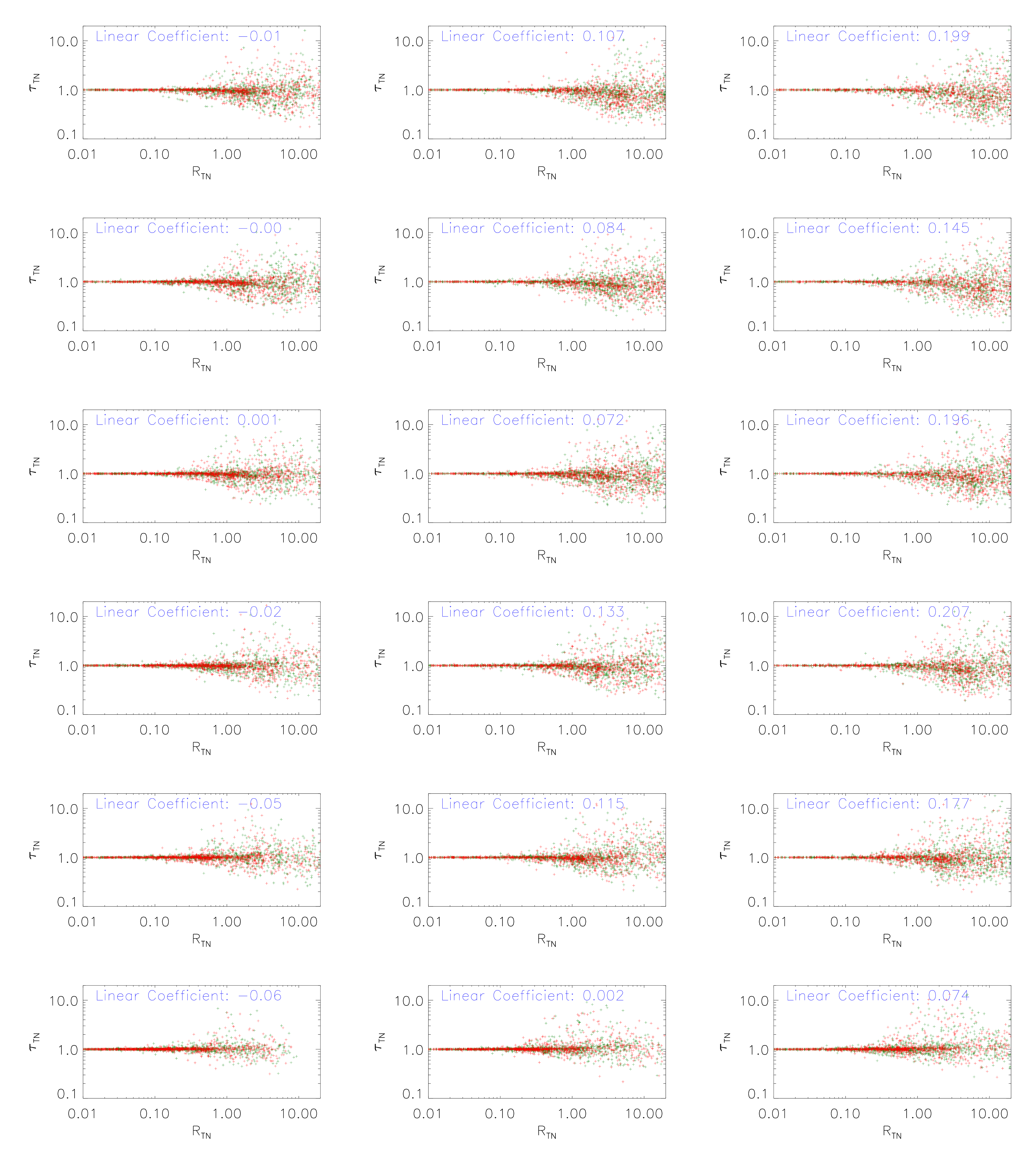}
\caption{On the dependence of $\tau_{TN}$ on $R_{TN}$ for the mock light curves based on the long-term variability  
$L_{\rm bol,~t,~CAR}$ created with $M_{BH}~=~10^6{\rm M_\odot}$, $\tau_0~=~200{\rm days}$ plus the $L_{\rm bol,~t,~TDE}$ created 
with $\gamma~=~4/3$. For the panels from top to bottom, the results are based on the redshift of 0.05, 0.1, 0.2, 0.3, 0.5, and 1.0, 
respectively. For the panels from left to right, the results are based on the $\eta$ of 0.06, 0.15 and 0.3, respectively. In each 
panel, the Spearman rank correlation coefficient for the correlation between $\tau_{TN}$ and $R_{TN}$ ($R_{TN}~>~1$) is listed in 
blue characters. In each panel, pluses in red and in dark green show the results with SNR larger than 55 and smaller than 55, 
respectively. In each panel, due to large number of dense data points, the error bars with uncertainties about 20\% - 25\% are 
not plotted.
}
\label{cd64}
\end{figure*}

\subsection{Results based on the Long-Term Variabilities of $L_{\rm bol,~t,~CAR}$}

	In the subsection, it is interesting to check effects of different model parameters on the dependence of $\tau_{TN}$ on 
$R_{TN}$ which are determined through the MPFIT package applied in the kbs09 method.

	Fig.~\ref{car} show two examples of the mock light curves $L_{\rm bol,~t}$ with different input model parameters marked 
in each panel. The first shown $L_{\rm bol,~t}$ is created with $\tau_0~=~200{\rm days}$, $M_{BH}~=~10^{6}{\rm M_\odot}$, $z~=~0.05$, 
$\gamma~=~4/3$, $\beta~\sim~1.55$, $M_*~\sim~0.31{\rm M_\odot}$, $T_{vis}~\sim~0.218$, $t_s~\sim~2200{\rm days}$, SNR~=~46, 
$\eta~=~0.06$. The second shown $L_{\rm bol,~t}$ is created with $\tau_0~=~600{\rm days}$, $M_{BH}~=~5\times10^{7}{\rm M_\odot}$, 
$z~=~1.0$, $\gamma~=~5/3$, $\beta~\sim~1.32$, $M_*~\sim~8.32{\rm M_\odot}$, $T_{vis}~\sim~0.085$, $t_s~\sim~1500{\rm days}$, 
SNR~=~30, $\eta~=~0.3$. Before proceeding further, there is a more intuitive result that smaller variability timescale of TDEs 
variability from cases with smaller BH mass should have few effects on $R_{TN}$, such as the shown results with tiny changes in 
timescales in top panels of Fig.~\ref{car}. More detailed results are shown as follows.

	Fig.~\ref{cd64} shows the dependence of $\tau_{TN}$ on $R_{TN}$ for the cases (cases-6-2-4, the first number '6' means 
BH mass as $10^6{\rm M_\odot}$, the second number '2' means $\tau_0/100{\rm days}~=~2$, and the third number '4' means 
$\gamma~\times~3=4$) with $M_{BH}~=~10^{6}{\rm M_\odot}$, $\tau_0~=~200{\rm days}$, and $\gamma~=~4/3$. It is clear that TDEs 
contributions around BHs with masses around $10^{6}{\rm M_\odot}$ have few effects on the dependence of $\tau_{TN}$ on $R_{TN}$, 
all results shown in Fig.~\ref{cd64} with Spearman rank correlation coefficients smaller than 0.3 for the data points with 
$R_{TN}~>~1$, even considering different redshift and different $\eta$. The results can be well expected due to smaller 
variability timescales of TDEs around BHs with masses around $10^{6}{\rm M_\odot}$, relative to the long time durations of 
$L_{\rm bol,~t,~CAR}$. Besides the results for the cases-6-2-4, there are totally similar results, no apparent positive 
dependence of $\tau_{TN}$ on $R_{TN}$, for the cases (cases-6-6-4) with $M_{BH}~=~10^{6}{\rm M_\odot}$, $\tau_0~=~600{\rm days}$, 
and $\gamma~=~4/3$, and for the cases (cases-6-2-5) with $M_{BH}~=~10^{6}{\rm M_\odot}$, $\tau_0~=~200{\rm days}$, and 
$\gamma~=~5/3$, and for the cases (cases-6-6-5) with $M_{BH}~=~10^{6}{\rm M_\odot}$, $\tau_0~=~600{\rm days}$, and $\gamma~=~5/3$. 
Therefore, we do not show the results on cases-6-6-4, cases-6-2-5, and cases-6-6-5\ in plots. And there are no further 
discussions on the results with $M_{BH}~=~10^{6}{\rm M_\odot}$, but the determined Spearman rank correlation coefficients 
are listed in Table~2 for all the cases with BH mass $10^6{\rm M_\odot}$. In one word, contributions of TDEs around BHs with 
masses $10^{6}{\rm M_\odot}$ cannot provide clear clues on central TDEs, through long-term variability.

	Then, similar as the discussed results on dependence of $\tau_{TN}$ on $R_{TN}$ for the cases with $M_{BH}=10^6{\rm M_\odot}$, 
the results on the dependence of $\tau_{TN}$ on $R_{TN}$ are also discussed with BH masses as $10^7{\rm M_\odot}$ and 
$5\times10^7{\rm M_\odot}$. Based on two different values of $M_{BH}$, two different values of $\tau_0$ and two different values 
of $\gamma$, there are 8 cases named as cases-7-2-4 (the first number '7' means BH mass as $\log(M_{BH}/{\rm M_\odot})=7$, the 
second number '2' means $\tau_0/100{\rm days}~=~2$, and the third number '4' means $\gamma~\times~3~=~4$), cases-7-6-4, 
cases-7-2-5, cases-7-6-5, cases-7.7-2-4 (the first number '7.7' means BH mass as $\log(M_{BH}/{\rm M_\odot})=\log(5\times10^7)\sim7.7$), 
cases-7.7-6-4, cases-7.7-2-5, cases-7.7-6-5. Then, similar as the discussed results for the $18\times4$ dependences for the four 
cases with $M_{BH}=10^6{\rm M_\odot}$, all the 144 (18$\times$8) dependences of $\tau_{TN}$ on $R_{TN}$ for $R_{TN}~>~R_{cri}$ are 
carefully checked in all the cases with $M_{BH}=10^7{\rm M_\odot}, 5\times10^7{\rm M_\odot}$. Here, the critical values 
$R_{cri}=0.3$ and $R_{cri}=0.15$ are simply determined and accepted for the cases with $M_{BH}=10^7{\rm M_\odot}$ and with 
$M_{BH}=5\times10^7{\rm M_\odot}$, respectively, after simply considering the variance of $\tau_{TN}$ of the data points with 
$R_{TN}~>~R_{cri}$ at least 2000 times larger than the variance of $\tau_{TN}$ of the data points with $R_{TN}~<~R_{cri}$. The 
determined Spearman Rank Correlation coefficients are listed in Table~2. Meanwhile, for the correlations with correlation 
coefficients larger than 0.3, through the same FITEXY code, the strong positive correlations between $\tau_{TN}$ and $R_{TN}$ 
for $R_{TN}~>~0.3$ can be well described by
\begin{equation}
	\log(\tau_{TN})~=~A~+~B~\times~\log(R_{TN})
\end{equation}
with determined $B$ also listed in Table~2.

\begin{figure*}
\centering\includegraphics[width = 18cm,height=20cm]{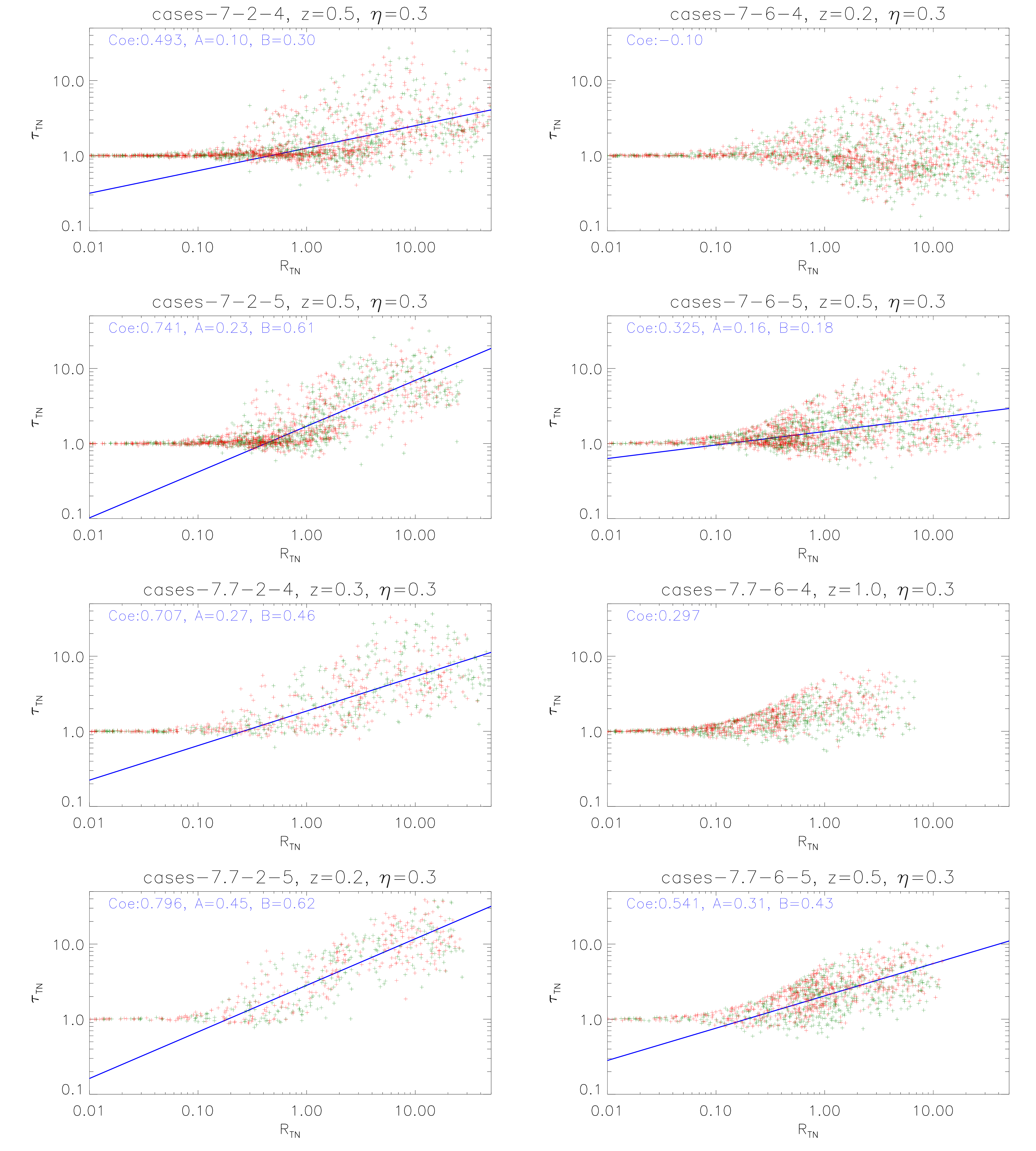}
\caption{On the dependences of $\tau_{TN}$ on $R_{TN}$ for the mock light curves based on the long-term variability 
$L_{\rm bol,~t,~CAR}$. The results in the eight panels show the dependence with maximum Spearman Rank correlation coefficient 
among the 18 dependences in cases-7-2-4 (with $M_{BH}=10^7{\rm M_\odot}$, $\tau_0=200days$ and $3\times\gamma=4$), cases-7-6-4 
(with $M_{BH}=10^7{\rm M_\odot}$, $\tau_0=600days$ and $3\times\gamma=4$), cases-7-2-5 (with $M_{BH}=10^7{\rm M_\odot}$, 
$\tau_0=200days$ and $3\times\gamma=5$), cases-7-6-5 (with $M_{BH}=10^7{\rm M_\odot}$, $\tau_0=600days$ and $3\times\gamma=5$), 
cases-7.7-2-4 (with $M_{BH}=5\times10^7{\rm M_\odot}$, $\tau_0=200days$ and $3\times\gamma=4$), cases-7.7-6-4 (with 
$M_{BH}=5\times10^7{\rm M_\odot}$, $\tau_0=600days$ and $3\times\gamma=4$), cases-7.7-2-5 (with $M_{BH}=5\times10^7{\rm M_\odot}$, 
$\tau_0=200days$ and $3\times\gamma=5$), cases-7.7-6-5 (with $M_{BH}=5\times10^7{\rm M_\odot}$, $\tau_0=600days$ and 
$3\times\gamma=5$) as the listed information in title of each panel. Meanwhile, the information of $z$ and $\eta$ are also listed in 
title of each panel. In each panel, Similar as the results shown in In each panel, pluses in red and in dark green show the results 
with SNR larger than 55 and smaller than 55, respectively. In each panel, due to large number of dense data points, the error 
bars with uncertainties about 20\% - 25\% are not plotted. In each panel, the calculated correlation coefficient (Coe) is marked 
in blue characters in top-left region for the correlation between $\tau_{TN}$ and $R_{TN}$ ($R_{TN}~>~R_{cri}$). In each panel with 
correlation coefficient larger than 0.3, solid blue line shows the linear description $\log(\tau_{TN})~=~A~+~B\log(R_{TN})$ to 
the correlation between $\tau_{TN}$ and $R_{TN}$ ($R_{TN}~>~R_{cri}$) and the determined parameters of $A$ and $B$ are listed 
in blue characters in top-left region. }
\label{cd775}
\end{figure*}

\begin{figure*}
\centering\includegraphics[width = 18cm,height=12cm]{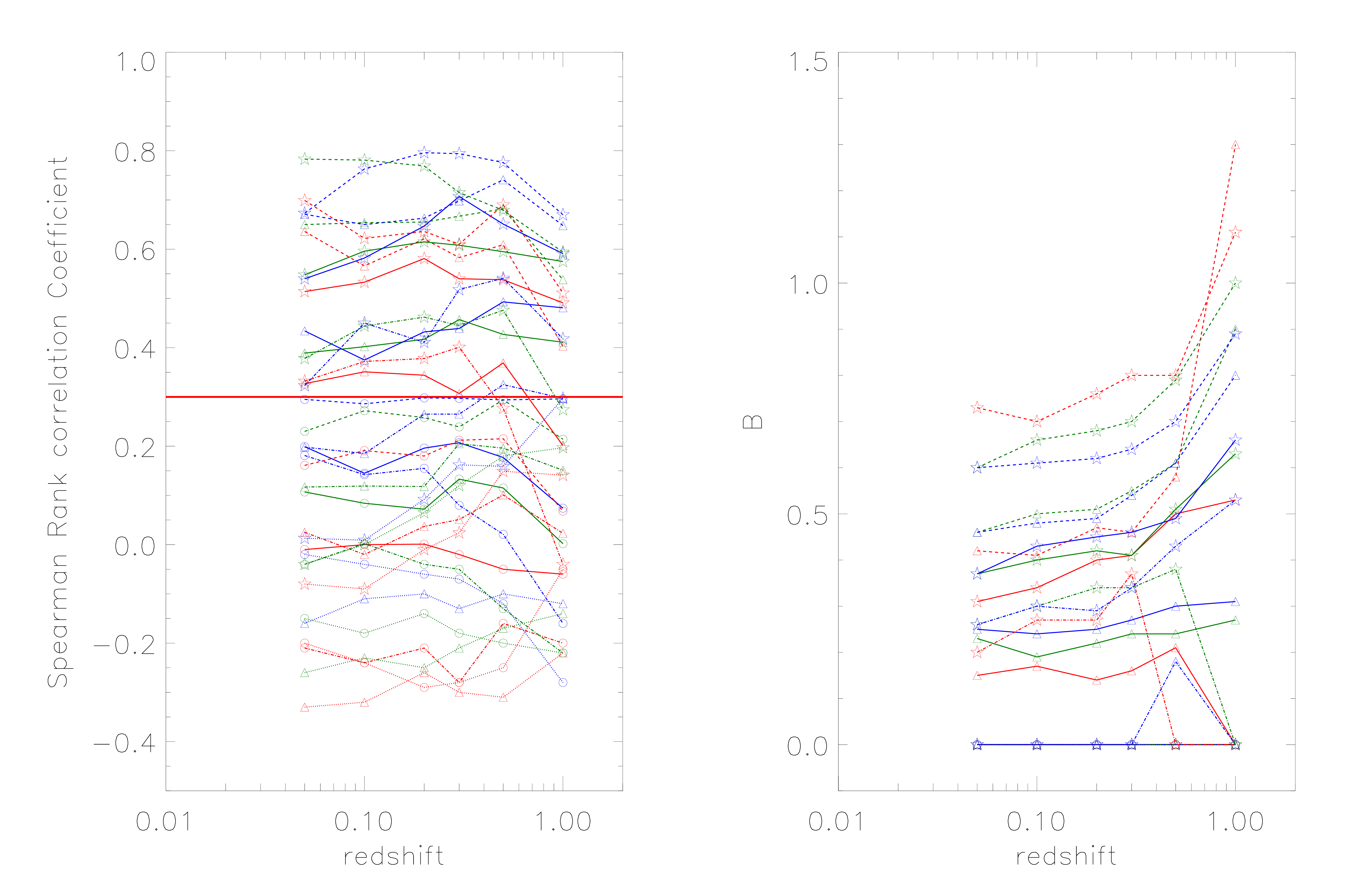}
\centering\includegraphics[width = 18cm,height=3cm]{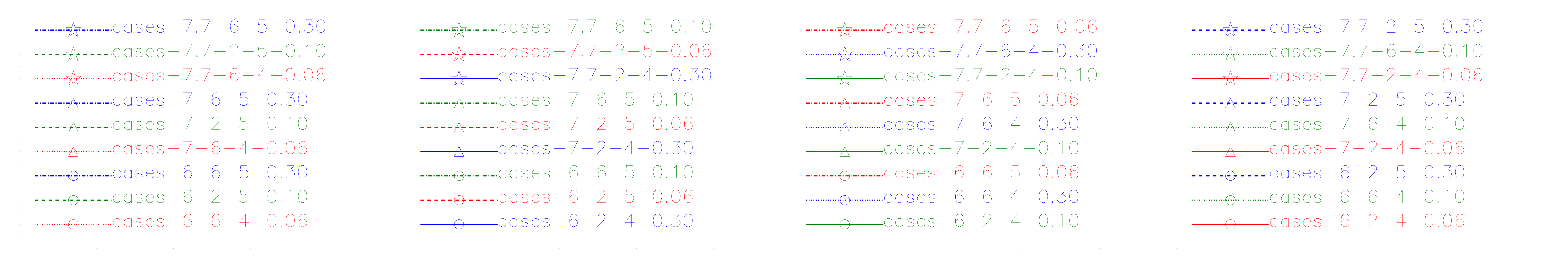}
\caption{Top left panel shows properties of Spearman Rank Correlation Coefficients for all the dependences of 
$\tau_{TN}$ on $R_{TN}$ for the cases with different $M_{BH}$, different $\tau_0$, different $z$, different $\gamma$ and 
different $\eta$. Top right panel shows properties of $B$ in the formula $\log(\tau_{TN})~=~A~+~B~\times~\log(R_{TN})$ for all 
the dependences. Bottom panel shows the legends used in top panels. The four numbers included in 'cases-n0-n1-n2-n3' shown in 
legends have the following meanings, 'n0' means logarithmic BH mass, 'n1' means the value of $\tau_0/100$, 'n2' means the 
values of $3\times\gamma$ and 'n2' means the value of $\eta$, for example, 'cases-7.7-2-4-0.30' means the 6 dependences (relative 
to six different values of redshift) of $\tau_{TN}$ on $R_{TN}$ for the case with $M_{BH}=5\times10^7{\rm M_\odot}$, 
$\tau_0=200{\rm days}$, $3\times\gamma=4/3$ and $\eta=0.30$. In top right panel, due to many dependences with coefficients 
smaller than 0.3, there are some dependences with their $B$ over-plotted with $B=0$. In top left panel, horizontal red line 
marks the position of Spearman Rank correlation coefficient of 0.3.}
\label{Adep}
\end{figure*}

        Here, not all the 144 (18$\times$8) dependences of $\tau_{TN}$ on $R_{TN}$ are shown in plots, but the dependence with 
maximum Spearman Rank correlation coefficient is shown in Fig.~\ref{cd775} among the 18 dependences in each case with 
$M_{BH}~=~10^{7}{\rm M_\odot}$, $M_{BH}~=~5~\times~10^{7}{\rm M_\odot}$. Meanwhile, based on the determined Coefficients and 
the slope $B$ (if there was) listed in Table~2 for the 216 dependences in the 12 cases with $M_{BH}~=~10^{6}{\rm M_\odot}$, 
$M_{BH}~=~10^{7}{\rm M_\odot}$, $M_{BH}~=~5~\times~10^{7}{\rm M_\odot}$, properties of Coefficients and slope $B$ are shown in 
Fig.~\ref{Adep}.

	Based on the determined Coefficients listed in Table~2 and the shown results in Fig.~\ref{Adep}, the following seven 
points can be found. First, comparing with the cases with $M_{BH}=10^{6}{\rm M_\odot}$, there are more sensitive and clearer 
positive dependence of $\tau_{TN}$ on $R_{TN}$ ($R_{TN}~>~0.3$), due to the results with Spearman rank correlation coefficients 
larger than 0.3: almost all the cases with input $\tau_0~=~200days$ and $M_{BH}~=~10^{7}{\rm M_\odot}$ have coefficients larger 
than 0.3 for the correlations with $R_{TN}~>~0.3$. Second, for the cases with $M_{BH}=10^{7}{\rm M_\odot}$, intrinsic variability 
timescales long as 600days should lead to no clear positive dependence of $\tau_{TN}$ on $R_{TN}$, but intrinsic variability 
timescales long as 200days can lead to clear positive dependence of $\tau_{TN}$ on $R_{TN}$. Third, for the cases with 
$M_{BH}~=~10^{7}{\rm M_\odot}$, the positive dependence of $\tau_{TN}$ on $R_{TN}$ are steeper (larger $B$) in the cases with 
$\gamma~=~5/3$ than with $\gamma~=~4/3$. Fourth, comparing with cases with $M_{BH}=10^{6}{\rm M_\odot}$ and 
$M_{BH}=10^{7}{\rm M_\odot}$, there are more sensitive and clearer positive dependence of $\tau_{TN}$ on $R_{TN}$ for the cases 
with $M_{BH}=5~\times~10^{7}{\rm M_\odot}$, due to the results with Spearman rank correlation coefficients larger than 0.3: all 
the cases with input $\tau_0~=~200days$ and half of the cases with $\tau_0~=~600days$ have coefficients larger than 0.3 for the 
correlation with $R_{TN}~>~0.15$. Fifth, for the cases with BH masses about $5~\times~10^{7}{\rm M_\odot}$, intrinsic variability 
timescales long as 600days but only with $\gamma~=~5/3$ should lead to clear positive dependence of $\tau_{TN}$ on $R_{TN}$, 
but intrinsic variability timescales long as 200days almost can lead to clear positive dependence of $\tau_{TN}$ on $R_{TN}$. 
Sixth, the positive dependence of $\tau_{TN}$ on $R_{TN}$ are steeper (larger $B$) in the cases with $\gamma~=~5/3$ than with 
$\gamma~=~4/3$. Seventh, there are few effects of SNR on dependences of $\tau_{TN}$ on $R_{TN}$, such as the shown results in 
Fig.~\ref{cd775}.

\begin{figure*}
\centering\includegraphics[width = 18cm,height=11cm]{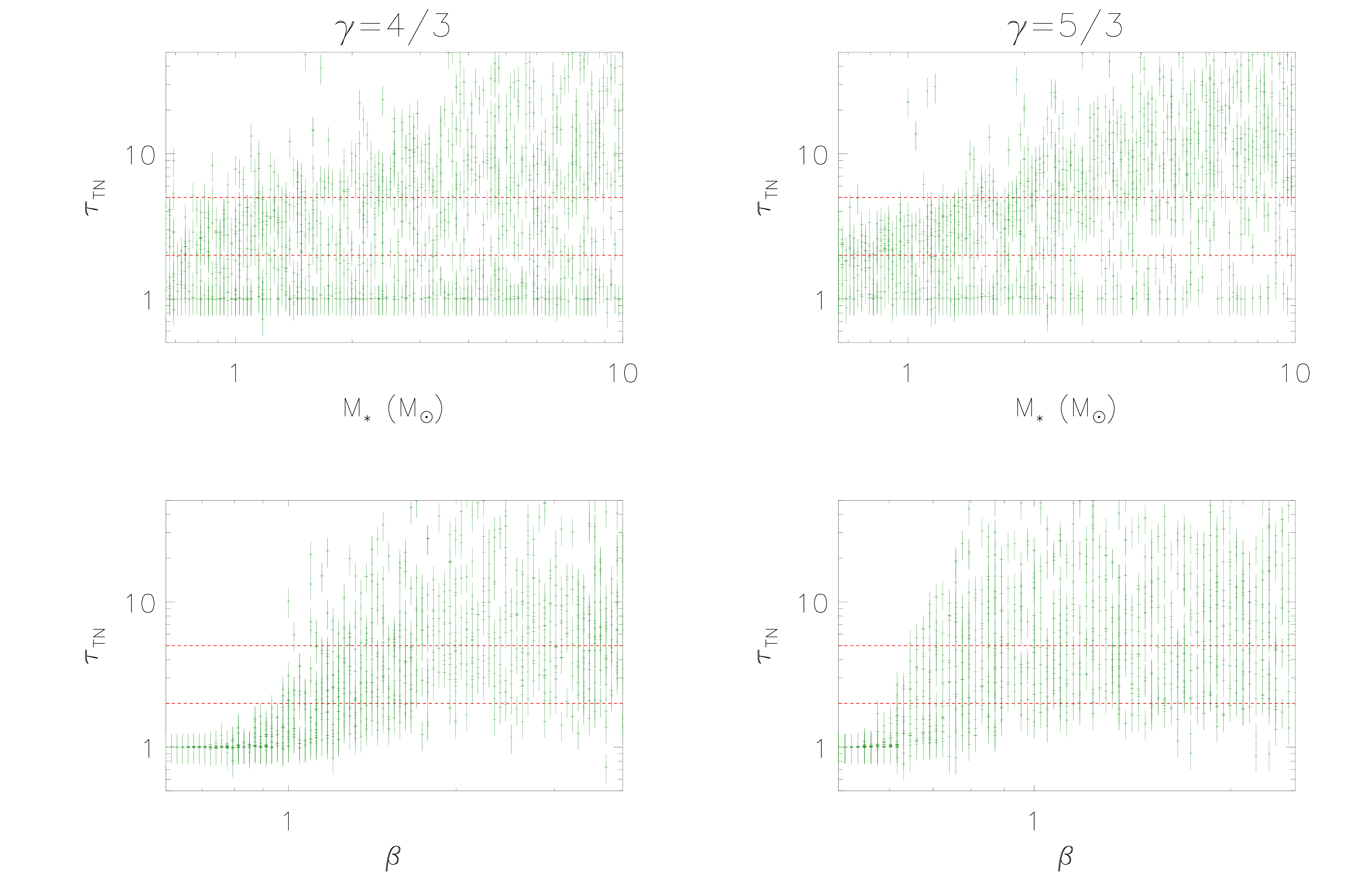}
\caption{Dependence of $\tau_{TN}$ on the stellar mass $M_\star$ (top panels) and on the impact parameter $\beta$ (bottom 
panels) for the simulated light curves $L_{\rm bol,~t}$ by $L_{\rm bol,~t,~N5548}$ plus TDEs contributions with $\gamma~=~4/3$ 
(left panels) and with $\gamma~=~5/3$ (right panels). In each panel, horizontal dashed red lines show $\tau_{TN}~=~2$ and 
$\tau_{TN}~=~5$, respectively.
}
\label{ms}
\end{figure*}

	Based on the results above, we can find that
\begin{itemize}
\item BH mass has apparent effects on the dependence of $\tau_{TN}$ on $R_{TN}$. Larger BH masses can lead to more apparent 
	and steeper dependence of $\tau_{TN}$ on $R_{TN}$.
\item Polytropic index $\gamma$ has apparent effects on the dependence of $\tau_{TN}$ on $R_{TN}$. $\gamma=5/3$ can lead to 
	more apparent and steeper dependence of $\tau_{TN}$ on $R_{TN}$.
\item Redshift has tiny effects on the dependence of $\tau_{TN}$ on $R_{TN}$. At least, redshift changing from 0.05 to 1.0 
	cannot lead to clear changes in the dependence of $\tau_{TN}$ on $R_{TN}$, only parameter $B$ being increased quite 
	smoothly in cases-7-2-5 (with $M_{BH}=10^7{\rm M_\odot}$, $\tau_0=200days$ and $3\times\gamma=5$), 
	cases-7.7-2-4 (with $M_{BH}=5\times10^7{\rm M_\odot}$, $\tau_0=200days$ and $3\times\gamma=4$), cases-7.7-2-5 (with 
	$M_{BH}=5\times10^7{\rm M_\odot}$, $\tau_0=200days$ and $3\times\gamma=5$) and cases-7.7-6-5 
	($M_{BH}=5\times10^7{\rm M_\odot}$, $\tau_0=600days$ and $3\times\gamma=5$).
\item Energy transfer efficiency has tiny effects on the dependence of $\tau_{TN}$ on $R_{TN}$. At least, energy transfer 
	efficiency changing from 0.06 to 0.3 cannot lead to clear changes in the dependence of $\tau_{TN}$ on $R_{TN}$.
\end{itemize}

\begin{table*}
\caption{Parameters $B$ and Spearman rank correlation coefficients for the dependence of $\tau_{TN}$ on $R_{TN}$}
\begin{center}
\begin{tabular}{cccccc|cccccc}
\hline\hline
$\tau_0$ & $\eta$ & $z$ & $M_6~=~1$ & $M_6~=~10$ & $M_6~=~50$ & 
	$\tau_0$ & $\eta$ & $z$ & $M_6~=~1$ & $M_6~=~10$ & $M_6~=~50$ \\
         &        &     & $B\ \ \ (\alpha)$ & $B\ \ \ (\alpha)$ & $B\ \ \ (\alpha)$ & 
	 &        &     & $B\ \ \ (\alpha)$ & $B\ \ \ (\alpha)$ & $B\ \ \ (\alpha)$ \\
\hline
\multicolumn{12}{c}{$\gamma~=~4/3$} \\
\hline
200 & 0.06 & 0.05 & 0~~ (-0.01) & 0.15~~ (0.327) & 0.31~~ (0.514) & 
	600 & 0.06 & 0.05 & 0~~ (-0.20) & 0~~ (-0.33) & 0~~ (-0.08)  \\
200 & 0.06 & 0.1 & 0~~ (-0.00) & 0.17~~ (0.351) & 0.34~~ (0.533) & 
	600 & 0.06 & 0.1 & 0~~ (-0.24) & 0~~ (-0.32) & 0~~ (-0.09)  \\
200 & 0.06 & 0.2 & 0~~ (0.001) & 0.14~~ (0.344) & 0.40~~ (0.581) & 
	600 & 0.06 & 0.2 & 0~~ (-0.29) & 0~~ (-0.26) & 0~~ (-0.01)  \\
200 & 0.06 & 0.3 & 0~~ (-0.02) & 0.16~~ (0.307) & 0.41~~ (0.540) & 
	600 & 0.06 & 0.3 & 0~~ (-0.28) & 0~~ (-0.30) & 0~~ (0.024)  \\
200 & 0.06 & 0.5 & 0~~ (-0.05) & 0.21~~ (0.369) & 0.50~~ (0.538) & 
	600 & 0.06 & 0.5 & 0~~ (-0.25) & 0~~ (-0.31) & 0~~ (0.149) \\
200 & 0.06 & 1.0 & 0~~ (-0.06) & 0~~ (0.201) & 0.53~~ (0.491) & 
	600 & 0.06 & 1.0 & 0~~ (-0.05) & 0 (-0.22) &  0~~ (0.141) \\
200 & 0.15 & 0.05 & 0~~ (0.107) & 0.23~~ (0.389) & 0.37~~ (0.548) & 
	600 & 0.15 & 0.05 & 0~~ (-0.15) & 0 (-0.26) & 0~~ (-0.04)  \\
200 & 0.15 & 0.1 & 0~~ (0.084) & 0.19~~ (0.402) & 0.40~~ (0.596) & 
	600 & 0.15 & 0.1 & 0~~ (-0.18) & 0~~ (-0.23) & 0~~ (-0.00)  \\
200 & 0.15 & 0.2 & 0~~ (0.072) & 0.22~~ (0.417) & 0.42~~ (0.615) & 
	600 & 0.15 & 0.2 & 0~~ (-0.14) & 0~~ (-0.25) & 0~~ (0.065)  \\
200 & 0.15 & 0.3 & 0~~ (0.133) & 0.24~~ (0.457) & 0.41~~ (0.608) & 
	600 & 0.15 & 0.3 & 0~~ (-0.18) & 0~~ (-0.21) & 0~~ (0.120)  \\
200 & 0.15 & 0.5 & 0~~ (0.115) & 0.24~~ (0.427) & 0.51~~ (0.595) & 
	600 & 0.15 & 0.5 & 0~~ (-0.20) & 0~~ (-0.17) & 0~~ (0.180)  \\
200 & 0.15 & 1.0 & 0~~ (0.002) & 0.27~~ (0.411) & 0.63~~ (0.575) & 
	600 & 0.15 & 1.0 & 0~~ (-0.22) & 0~~ (-0.14) & 0~~ (0.197)  \\
200 & 0.3 & 0.05 & 0~~ (0.199) & 0.25~~ (0.434) & 0.37~~ (0.540) & 
	600 & 0.3 & 0.05 & 0~~ (-0.02) & 0~~ (-0.16) & 0~~ (0.013)  \\
200 & 0.3 & 0.1 & 0~~ (0.145) & 0.24~~ (0.375) & 0.43~~ (0.582) & 
	600 & 0.3 & 0.1 & 0~~ (-0.04) & 0~~ (-0.11) & 0~~ (0.009)  \\
200 & 0.3 & 0.2 & 0~~ (0.196) & 0.25~~ (0.432) & 0.45~~ (0.647) & 
	600 & 0.3 & 0.2 & 0~~ (-0.06) & 0~~ (-0.10) & 0~~ (0.092)  \\
200 & 0.3 & 0.3 & 0~~ (0.207) & 0.27~~ (0.439) & 0.46~~ (0.707) & 
	600 & 0.3 & 0.3 & 0~~ (-0.07) & 0~~ (-0.13) & 0~~ (0.162)  \\
200 & 0.3 & 0.5 & 0~~ (0.177) & 0.30~~ (0.493) & 0.49~~ (0.651) & 
	600 & 0.3 & 0.5 & 0~~ (-0.12) & 0~~ (-0.10) & 0~~ (0.159)  \\
200 & 0.3 & 1.0 & 0~~ (0.074) & 0.31~~ (0.481) & 0.66~~ (0.591) & 
	600 & 0.3 & 1.0 & 0~~ (-0.28) & 0~~ (-0.12) & 0~~ (0.297) \\
\hline
\multicolumn{12}{c}{$\gamma~=~5/3$} \\
\hline
200 & 0.06 & 0.05 & 0~~ (0.161) & 0.42~~ (0.636) & 0.73~~ (0.699) & 
	600 & 0.06 & 0.05 & 0~~ (-0.21) & 0~~ (0.025) & 0.20~~ (0.332)  \\
200 & 0.06 & 0.1 & 0~~ (0.191) & 0.41~~ (0.566) & 0.70~~ (0.622) & 
	600 & 0.06 & 0.1 & 0~~ (-0.24) & 0~~ (-0.02) & 0.27~~ (0.372)  \\
200 & 0.06 & 0.2 & 0~~ (0.180) & 0.47~~ (0.621) & 0.76~~ (0.636) & 
	600 & 0.06 & 0.2 & 0~~ (-0.21) & 0~~ (0.037) & 0.27~~ (0.378)  \\
200 & 0.06 & 0.3 & 0~~ (0.212) & 0.46~~ (0.583) & 0.80~~ (0.610) & 
	600 & 0.06 & 0.3 & 0~~ (-0.28) & 0~~ (0.051) & 0.37~~ (0.401)  \\
200 & 0.06 & 0.5 & 0~~ (0.215) & 0.58~~ (0.609) & 0.80~~ (0.690) & 
	600 & 0.06 & 0.5 & 0~~ (-0.16) & 0~~ (0.101) & 0~~ (0.277)  \\
200 & 0.06 & 1.0 & 0~~ (0.068) & 1.30~~ (0.403) & 1.11~~ (0.511) & 
	600 & 0.06 & 1.0 & 0~~ (-0.20) & 0~~ (0.023) & 0~~ (-0.04)  \\
200 & 0.15 & 0.05 & 0~~ (0.230) & 0.46~~ (0.650) & 0.60~~ (0.783) & 
	600 & 0.15 & 0.05 & 0~~ (-0.04) & 0~~ (0.117) & 0.26~~ (0.378)  \\
200 & 0.15 & 0.1 & 0~~ (0.272) & 0.50~~ (0.654) & 0.66~~ (0.781) & 
	600 & 0.15 & 0.1 & 0~~ (0.002) & 0~~ (0.119) & 0.30~~ (0.444)  \\
200 & 0.15 & 0.2 & 0~~ (0.258) & 0.51~~ (0.655) & 0.68~~ (0.769) & 
	600 & 0.15 & 0.2 & 0~~ (-0.04) & 0~~ (0.118) & 0.34~~ (0.462)  \\
200 & 0.15 & 0.3 & 0~~ (0.239) & 0.55~~ (0.667) & 0.70~~ (0.715) & 
	600 & 0.15 & 0.3 & 0~~ (-0.05) & 0~~ (0.205) & 0.34~~ (0.445)  \\
200 & 0.15 & 0.5 & 0~~ (0.294) & 0.61~~ (0.682) & 0.79~~ (0.679) & 
	600 & 0.15 & 0.5 & 0~~ (-0.13) & 0~~ (0.196) & 0.38~~ (0.476)  \\
200 & 0.15 & 1.0 & 0~~ (0.214) & 0.90~~ (0.538) & 1.00~~ (0.594) & 
	600 & 0.15 & 1.0 & 0~~ (-0.22) & 0~~ (0.151) & 0~~ (0.274)  \\
200 & 0.3 & 0.05 & 0~~ (0.295) & 0.46~~ (0.671) & 0.60~~ (0.673) & 
	600 & 0.3 & 0.05 & 0~~ (0.181) & 0~~ (0.198) & 0.26~~ (0.323)  \\
200 & 0.3 & 0.1 & 0~~ (0.286) & 0.48~~ (0.650) & 0.61~~ (0.763) & 
	600 & 0.3 & 0.1 & 0~~ (0.142) & 0~~ (0.185) & 0.30~~ (0.450)  \\
200 & 0.3 & 0.2 & 0~~ (0.298) & 0.49~~ (0.663) & 0.62~~ (0.796) & 
	600 & 0.3 & 0.2 & 0~~ (0.155) & 0~~ (0.265) & 0.29~~ (0.411)  \\
200 & 0.3 & 0.3 & 0~~ (0.297) & 0.54~~ (0.698) & 0.64~~ (0.794) & 
	600 & 0.3 & 0.3 & 0~~ (0.080) & 0~~ (0.265) & 0.34~~ (0.518)  \\
200 & 0.3 & 0.5 & 0~~ (0.294) & 0.61~~ (0.741) & 0.70~~ (0.776) & 
	600 & 0.3 & 0.5 & 0~~ (0.021) & 0.18~~ (0.325) &  0.43~~ (0.541) \\
200 & 0.3 & 1.0 & 0~~ (0.296) & 0.80~~ (0.648) & 0.89~~ (0.670) & 
	600 & 0.3 & 1.0 & 0~~ (-0.16) & 0~~ (0.297) & 0.53~~ (0.418) \\
\hline\hline
\end{tabular} \\
\end{center}
\begin{itemize}
\item[1: ]From column 4 to column 6 and from column 10 to column 12, there are two numbers '$B\ \ \ (\alpha)$' in each cell, 
where $B$ is the slope of the formula $\log(\tau_{TN})~=~A~+~B~\times~\log(R_{TN})$, and $\alpha$ is the determined Spearman rank 
correlation coefficient for the data points with $R_{TN}$ larger than the critical values shown in textbody. If $\alpha$ is 
smaller than 0.3, then $B$ is set to be zero. 
\item[2: ]$M_6$ means the BH mass in unit of $10^6{\rm M_\odot}$. 
\item[3: ]$\tau_0$ means the input variability timescale in unit of days, when the third kind of $L_{\rm bol,~t}$ are simulated.
\end{itemize}
\end{table*}

\section{Discussions and further applications}

	It is necessary to check whether intrinsic AGN variability can provide quite different variability timescales in 
different epochs. Here, based on the 13years-long light curve $L_{\rm bol,~t,~N5548}$, 100 different 2000days-long (about 10times 
of the intrinsic variability timescale 200days) light curves can be randomly collected from $L_{\rm bol,~t,~N5548}$ with time 
duration from a randomly given starting time $0~<~t_0/{\rm days}~<~3600$ to $t_0~+~2000$. The kbs09 method is applied to determine 
the variability timescales $\tau_d$ of the 100 different 2000days-long light curves. Then, we can find that the ratios of $\tau_d$ 
to the variability timescale 219days of $L_{\rm bol,~t,~N5548}$ have mean value 1.02 with standard deviation 0.11. It is clear 
that light curves in different epochs cannot lead variability timescale varying so large as the results shown in Fig.~\ref{ns} 
with large TDEs contributions. Similar results can be found from the mock light curves of $L_{\rm bol,~t,~AGN}$ and 
$L_{\rm bol,~t,~CAR}$.

	Furthermore, there are seven more points we should note. First, in order to find more clearer effects of TDEs contributions 
on long-term AGN variability, the time duration is longer as 13years in the $L_{\rm bol,~t}$. Once there were shorter time 
durations applied, the dependence of $\tau_{TN}$ on $R_{TN}$ would have larger scatters, due to probably only part of TDEs 
contributions covered in $L_{\rm bol,~t}$. Moreover, the simulating light curves are based on BH masses smaller than 
$10^8{\rm M_\odot}$. When BH mass is larger than $10^8{\rm M_\odot}$, more massive but shorter-lived main-sequence stars are 
necessary to simulate suitable TDEs, otherwise tidal disruption radius should be smaller than event horizon of central BH. 
Therefore, the large BH mass is selected to be $5\times10^7~{\rm M_\odot}$ in the manuscript.

	Second, as the discussed and shown results in \citet{mi10, kbs09}, the parameter $\sigma$ and $\tau$ are probably connected. 
However, the connection between $\sigma$ and $\tau$ is quite loose. Therefore, in the manuscript, there are no considerations of 
the connection $\sigma$ and $\tau$, when the third kind of mock light curves $L_{\rm bol,~t}$ are simulated. With the similar 
considerations, due to the loose dependence of energy transfer efficiency and BH mass discussed in \citet{dl11}, the energy 
transfer efficiency $\eta$ is randomly selected to be 0.06, 0.15 and 0.3. Otherwise, the expected energy transfer efficiency around 
$M_{BH}=10^6{\rm M_\odot}$ should be small to be 0.008, an extremely smaller value.

	Third, besides BH masses and intrinsic variability timescales, there are no further considerations on the other parameters 
related to TDEs model. Actually, the parameters, such as the stellar mass $M_\star$ and impact parameter $\beta$, should have 
effects on the $\tau_{TN}$, because commonly larger $M_\star$ and $\beta$ can commonly lead to stronger TDEs expected bolometric 
luminosities. As examples, Fig.~\ref{ms} shows the dependence of $\tau_{TN}$ on the stellar mass $M_\star$ and on the impact 
parameter $\beta$ for the simulated light curves $L_{\rm bol,~t}$ by $L_{\rm bol,~t,~N5548}$ plus TDEs contributions. For the 
shown dependence of $\tau_{TN}$ on the stellar mass $M_\star$, there are positive correlations with Spearman rank correlation 
coefficients about 0.35 and 0.61 ($P_{null}~<~10^{-15}$) for the results with $\gamma~=~4/3$ and with $\gamma~=~5/3$, respectively. 
And, for the shown dependence of $\tau_{TN}$ on the $\beta$, there are positive correlations with Spearman rank correlation 
coefficients about 0.76 and about 0.54 ($P_{null}<10^{-15}$) for the results with $\gamma~=~4/3$ and with $\gamma~=~5/3$, 
respectively. Even for $M_\star$ around one solar mass or $\beta$ gently larger than 1, $\tau_{TN}$ can be well larger than 2. 
Certainly, for the cases with smaller BH masses, the positive correlations on $M_\star$ and on $\beta$ should be not so strong. 
However, not similar as the central BH masses and redshift of normal AGN which can be well estimated through spectroscopic features, 
the $M_\star$ and $\beta$ can not be previously measured. And the main objective is to provide clues to detect probable hidden 
TDEs in normal AGN. Probability of more massive main-sequence stars being tidally disrupted with larger $\beta$ in TDEs in normal 
AGN is not the objective of the manuscript. If there was a more massive main-sequence star was tidally disrupted with larger 
$\beta$ in a normal broad line AGN, it would be more preferred to detect the expected hidden TDEs. Therefore, in the manuscript, 
effects of the model parameters related to the theoretical TDEs model are not discussed.

	Fourth, as discussed in \citet{ks17}, shorter time baseline should lead to underestimated process parameter $\tau$ 
in DRW/CAR process. Considering the expected longer $\tau$ due to larger contributions from TDEs, intrinsic values of process 
parameter $\tau$ should be larger than the currently determined values for the created mock light curves. Therefore, combining 
with the input value of process parameter $\tau$ for $L_{\rm bol,~t,~CAR}$, larger values of $\tau_{TN}$ could be expected, 
leading to more apparent dependence of $\tau_{TN}$ on $R_{TN}$ to support our final conclusions. Meanwhile, accepted the criterion 
reported in \citet{ks17} that there are good estimations of process parameters for light curves with $\tau/t_{exp}<0.1$ (similar 
to process parameter $\tau$ divided by time baseline), the determined parameters are not biased for the mock light curves created 
with $\tau_0=200$days ($\tau/t_{exp}\sim{\rm 200days}/{\rm 13years}\sim0.04<0.1$). Therefore, even only considering the results 
based on $\tau_0=200$days, similar conclusions on effects of TDEs contributions can be given.

	Fifth, the standard theoretical TDE model discussed in \citet{gr13, gm14, mg19} is applied in the manuscript, leading 
to expected time-dependent decline $t^{-5/3}$ at late times. However, besides standard TDE model expected variability pattern, 
there are slow TDEs, such as the discussed results in \citet{gd17}, probably leading to shallower decline closer to $t^{-1}$. 
The slow TDEs could lead to much longer time durations than standard TDEs. However, based on the discussed results above, more 
apparent difference between characteristic timescales of TDEs variability and characteristic timescales of intrinsic AGN 
variability should lead to more apparent dependence of $\tau_{TN}$ on $R_{TN}$. Therefore, even without considering rare numbers 
of slow TDEs, considerations of slow TDEs could lead to more apparent clues to support our final conclusions.

	Sixth, as more recent discussions in \citet{bs22}, host galaxy dilution could have strong effects on determined process 
parameter. However, accepted host galaxy contribution as an constant component with none variability (almost inevitable), the 
host galaxy dilution should have few effects on process parameter of $\tau$, because the host galaxy contribution can be included 
in the parameter $L_{b0}$ in Equation (12) above. In the manuscript, the ratio of $\tau$ from the light curves with and without 
TDEs contributions are mainly considered, therefore, the host galaxy dilution has few effects on our final conclusions.

	Seventh, although all the quasars with measurements of continuum luminosities are collected from \citet{sh11} to 
determine the dependence of bolometric luminosity on redshift shown in Fig.~\ref{Bz}, some weak quasars are actually not 
included in the collected quasars, due to their lower continuum emissions. However, considering the very loose (or very weakly 
positive) dependence of DRW process parameter $\tau$ on luminosity as simply discussed in \citet{kbs09, mi10}, lower bolometric 
luminosities should lead to no variations of (or lower) DRW process parameter $\tau$ of intrinsic AGN variability. Therefore, 
even considering contributions of the lost weak quasars, there should be not different conclusions if accepted no dependence of 
DRW process parameter $\tau$ on luminosity, or lead to more apparent clues to support our final conclusions if accepted 
weakly positive dependence of DRW process parameter $\tau$ on luminosity.


        Based on the expected effects of TDEs contributions on long-term AGN variability, to check variability properties 
in different epochs of normal AGN could provide clues on probable central hidden TDEs in normal AGN with apparently intrinsic 
variability. In one word, the results in the manuscript provide an interesting and practical method to detect probable 
hidden TDEs in normal AGN with apparent intrinsic variability, especially for AGN with smaller intrinsic variability timescales 
but BH masses larger than $10^7{\rm M_\odot}$. To report detected hidden TDEs in normal AGN through quite different $\tau$ in 
different epochs can provide robust evidence to support the results in the manuscript. Considering TDEs expected time 
durations about several years for $M_{BH}\sim10^7{\rm M_\odot}$, baseline about (more than) 10years-long should be necessary for 
light curves to detected clues for hidden TDEs in broad line AGN. Therefore, combining light curves from different sky survey 
projects should be the efficient way to build light curves with baseline longer than 10 years. Unfortunately, there are quite 
different qualities, such as different baseline, different time steps, different SNRs, different covered wavelength ranges, 
different transmission curves for different filters, etc., for light curves from different sky survey projects. Before checking  
probably different intrinsic variability properties in different epochs from different sky survey projects, effects of the quite 
different qualities should be firstly and clearly determined. In current stage, long-term light curves from CSS and from ZTF 
for a large sample of SDSS quasars have been collected, and basic results are currently in writing. In the near future, effects 
of different qualities on variability properties for light curves from CSS and ZTF and a small sample of quasars with quite 
different $\tau$ in light curves from CSS and from ZTF will be discussed and reported as soon as possible. It is a 
great pity that we can not currently give a clear estimation on detection rates of hidden TDEs through combinations of light 
curves from different sky survey projects, especially because we do not know what key parameters related to AGN dominate  
probable TDEs contributions. However, the results in the manuscript are showing a practicable way to detect hidden AGN in 
normal broad line AGN with apparent variability. To detect hidden TDEs in broad line AGN through different variability properties 
in different epochs is our main objective in the near future.

\section{conclusions}
	Finally, we give our main conclusions as follows. Based on the AGN variability templates simulated by the CAR process 
and the variability from theoretical TDEs model, effects of TDEs contributions can be well estimated on the long-term variability 
properties of normal AGN with apparent intrinsic variability. Stronger TDEs contributions can lead to longer variability timescale 
$\tau$ of observational long-term AGN variability, especially for AGN with smaller intrinsic variability timescales and with BH 
masses larger than $10^7{\rm M_\odot}$. Therefore, the results provide an interesting forward-looking and practicable method to 
detect central hidden TDEs in normal broad line AGN based on quite different variability properties in different epochs, 
especially in broad line AGN with shorter intrinsic variability timescales and with BH masses larger than $10^7{\rm M_\odot}$.

\section*{Acknowledgements}
Zhang gratefully acknowledges the anonymous referee for giving us constructive comments and suggestions to greatly 
improve our paper. Zhang gratefully acknowledges the kind support from the Chinese grant NSFC-12173020 and NSFC-12373014. 
The paper has made use of the code of TDEFIT \url{https://tde.space/tdefit/} which is s a piece of open-source software written 
by James Guillochon for the purposes of model-fitting photometric light curves of tidal disruption events, and also made use of the 
code of MOSFIT (Modular Open Source Fitter for Transients) \url{https://mosfit.readthedocs.io/} which is a Python 2.7/3.x package 
for fitting, sharing, and estimating the parameters of transients via user-contributed transient models. The paper has made use 
of the data of NGC 5548 from AGNWATCH project (\url{https://www.asc.ohio-state.edu/astronomy/agnwatch/}) which is a consortium 
of astronomers who have studied the inner structure of AGN through continuum and emission-line variability. The paper has 
made use of the public code of FITEXY from the IDL Astronomy User's Library (\url{https://idlastro.gsfc.nasa.gov/ftp/pro/math/}). 
The paper has made use of the MCMC code \url{https://emcee.readthedocs.io/en/stable/index.html}, and mase use of the MPFIT package 
\url{https://pages.physics.wisc.edu/~craigm/idl/cmpfit.html}.

\section*{Data Availability}
The data underlying this article will be shared on reasonable request to the corresponding author
(\href{mailto:xgzhang@gxu.edu.cn}{xgzhang@gxu.edu.cn}).

\label{lastpage}

\begin{thebibliography}{   }
\bibitem[\protect\citeauthoryear{Anderson et al.}{2020}]{an20}
Anderson, M.; Mooley, K.; Hallinan, G.; et al., 2020, ApJ, 903, 116 – 127
\bibitem[\protect\citeauthoryear{Andrae, Kim \& Bailer-Jones}{2013}]{ak13}
Andrae R.; Kim, D. W.; Bailer-Jones C. A. L., 2013, A\&A, 554, 137
\bibitem[\protect\citeauthoryear{Bailer-Jones}{2012}]{bj12}
Bailer-Jones C. A. L., 2012, A\&A, 546, A89
\bibitem[\protect\citeauthoryear{Baldassare, Geha \& Greene}{2020}]{bv20}
Baldassare, V. F.; Geha, M.; Greene, J., 2020, ApJ, 896, 10
\bibitem[\protect\citeauthoryear{Bentz et al.}{2010}]{bw10}
Bentz, M. C.; Walsh, J. L.; Barth, A. J., et al., 2010, ApJ, 716, 993
\bibitem[\protect\citeauthoryear{Blagorodnova et al.}{2017}]{bl17}
Blagorodnova, N.; Gezari, S.; Hung, T.; et al., 2017, ApJ, 844, 46
\bibitem[\protect\citeauthoryear{Blanchard et al.}{2017}]{bn17}
Blanchard, P. K., Nicholl, M., Berger, E., et al., 2017, ApJ, 843, 106
\bibitem[\protect\citeauthoryear{Bonnerot et al.}{2016}]{br16}
Bonnerot, C.; Rossi, E. M.; Lodato, G.; Price, D. J., 2016, MNRAS, 455, 2253
\bibitem[\protect\citeauthoryear{Brockwell \& Davis}{2002}]{bd02}
Brockwell, P. J.; Davis, R. A. 2002, Introduction to Time Series and Forecasting (2nd ed.; New York: Springer)
\bibitem[\protect\citeauthoryear{Burke et al.}{2020}]{bs20}
Burke, C. J.; Shen, Y.; Chen, Y.; Scaringi, S.; Faucher-Giguere, C.; Liu, X.; Yang, Q, 2020, ApJ, 899, 136
\bibitem[\protect\citeauthoryear{Burke et al.}{2021}]{bs21}
Burke, C. J.; Shen, Y.; Blaes, O., et al., 2021, Sci, 373, 789
\bibitem[\protect\citeauthoryear{Burke et al.}{2022}]{bs22}
Burke, C. J.; Liu, X.; Shen, Y.; et al., 2022, MNRAS, 516, 2736
\bibitem[\protect\citeauthoryear{Cenko et al.}{2012}]{ce12}
Cenko S. B.; Krimm, H. A.; Horesh, A.; et al., 2012, ApJ, 753, 77
\bibitem[\protect\citeauthoryear{Chan et al.}{2019}]{cp19}
Chan, C.-H.; Piran, T.; Krolik, J. H.; Saban, D., 2019, ApJ, 881, 113
\bibitem[\protect\citeauthoryear{Chan, Piran \& Krolik}{2020}]{cp20}
Chan, C.-H.; Piran, T.; Krolik, J. H., 2020, ApJ, 903, 17
\bibitem[\protect\citeauthoryear{Chen, Dou \& Shen}{2022}]{cd22}
Chen, J.; Dou, L.; Shen, R., 2022, ApJ, 928, 63
\bibitem[\protect\citeauthoryear{Chornock et al.}{2014}]{ch14}
Chornock, R.; Berger, E.; Gezari, S.; et al., 2014, ApJ, 780, 44
\bibitem[\protect\citeauthoryear{Davis \& Laor}{2011}]{dl11}
Davis, S. W.; Laor, A., 2011, ApJ, 728, 98
\bibitem[\protect\citeauthoryear{Dexter \& Begelman}{2019}]{db19}
Dexter, J., Begelman, M. C., 2019, MNRAS Letter, 483, 17
\bibitem[\protect\citeauthoryear{Drake et al.}{2009}]{dr09}
Drake, A. J.; Djorgovski, S. G.; Mahabal, A.; et al., 2009, ApJ, 696, 870
\bibitem[\protect\citeauthoryear{Drake et al.}{2011}]{dd11}
Drake, A. J.; Djorgovski, S. G.; Mahabal, A.; et al., 2011, ApJ, 735, 106
\bibitem[\protect\citeauthoryear{Duras et al.}{2020}]{du20}
Duras, F.; Bongiorno, A.; Ricci, F.; et al., 2020, A\&A, 636, 73
\bibitem[\protect\citeauthoryear{Favre, Courevoisier \& Paltani}{2005}]{fc05}
Favre P.; Courevoisier T. J. L.; Paltani S., 2005, A\&A, 443, 451
\bibitem[\protect\citeauthoryear{Foreman-Mackey et al.}{2013}]{fh13}
Foreman-Mackey, D.; Hogg, D. W.; Lang, D.; Goodman, J., 2013, PASP, 125, 306
\bibitem[\protect\citeauthoryear{Gezari et al.}{2012}]{gs12}
Gezari S.; Chornock, R.; Rest, A.; et al., 2012, Nature, 485, 217
\bibitem[\protect\citeauthoryear{Gezari}{2021}]{gs21}
Gezari S., 2021, ARA\&A, 59, 21
\bibitem[\protect\citeauthoryear{Graham et al.}{2017}]{gd17}
Graham, M. J.; Djorgovski, S. G.; Drake, A. J. ; Stern, D.; Mahabal, A. A.; Glikman, E.; Larson, S.; 
		Christensen, E., 2017, MNRAS, 470, 4112
\bibitem[\protect\citeauthoryear{Gromadzki et al.}{2019}]{gr19}
Gromadzki, M.; Hamanowicz, A.; Wyrzykowski, L.; et al., 2019, A\&A, 622, 2
\bibitem[\protect\citeauthoryear{Guillochon \& Ramirez-Ruiz}{2013}]{gr13}
Guillochon, J.; \& Ramirez-Ruiz, E., 2013, ApJ, 767, 25
\bibitem[\protect\citeauthoryear{Guillochon, Manukian \& Ramirez-Ruiz}{2014}]{gm14}
Guillochon, J.; Manukian, H.; Ramirez-Ruiz, E., 2014, ApJ, 783, 23
\bibitem[\protect\citeauthoryear{Guillochon et al.}{2018}]{gn18}
Guillochon, J.; Nicholl, M.; Villar, A., et al., 2018, ApJS, 236, 6 
\bibitem[\protect\citeauthoryear{Guo et al.}{2017}]{gw17}
Guo, H.; Wang, J.; Cai, Z.; Sun, M., 2017, ApJ, 847, 132
\bibitem[\protect\citeauthoryear{Hawkins}{2002}]{ha02}
Hawkins, M. R. S., 2002, MNRAS, 329, 76
\bibitem[\protect\citeauthoryear{Hayasaki, Stone \& Loeb}{2016}]{hs16}
Hayasaki, K.; Stone, N.; Loeb, A., 2016, MNRAS, 461, 3760
\bibitem[\protect\citeauthoryear{Hinkle et al.}{2021}]{hi21}
Hinkle, J. T.; Holoien, T. W. -S.; Auchettl, K.; et al., 2021, MNRAS, 500, 1673
\bibitem[\protect\citeauthoryear{Holoien et al.}{2014}]{ht14}
Holoien, T. W.; Prieto, J. L.; Bersier, D.; et al., 2014, MNRAS, 445, 3263
\bibitem[\protect\citeauthoryear{Holoien et al.}{2016}]{ho16}
Holoien, T. W. S.; Kochanek, C. S.; Prieto, J. L.; et al., 2016, MNRAS, 455, 2918
\bibitem[\protect\citeauthoryear{Holoien et al.}{2019}]{ht19}
Holoien, T. W. S.; Huber, M. E.; Shappee, B. J.; et al., 2019, ApJ, 880, 120
\bibitem[\protect\citeauthoryear{Horne et al.}{2021}]{hd21}
Horne, K.; De Rosa, G.; Peterson, B. M., et al., 2021, ApJ, 907, 76
\bibitem[\protect\citeauthoryear{Hung et al.}{2020}]{hf20}
Hung, T.; Foley, R. J.; Ramirez-Ruiz, E.; et al., 2020, ApJ, 903, 31
\bibitem[\protect\citeauthoryear{Kasliwal et al.}{2015}]{kv15}
Kasliwal, V. P.; Vogeley, M. S.; Richards, G. T., 2015, MNRAS, 451, 4328
\bibitem[\protect\citeauthoryear{Kathirgamaraju et al.}{2017}]{kb17}
Kathirgamaraju, A.; Barniol Duran, R.; Giannios, D., et al., 2017, MNRAS, 469, 314
\bibitem[\protect\citeauthoryear{Kelly, Bechtold \& Siemiginowska}{2009}]{kbs09}
Kelly B. C.; Bechtold J.; Siemiginowska A., 2009, ApJ, 698, 895
\bibitem[\protect\citeauthoryear{Kelly et al.}{2014}]{kb14}
Kelly, B. C.; Becker, A. C.; Sobolewska, M.; Siemiginowska, A.; Uttley, P., 2014, ApJ, 788, 33
\bibitem[\protect\citeauthoryear{Kochanek}{1994}]{kc94}
Kochanek, C. S., 1994, ApJ, 422, 508
\bibitem[\protect\citeauthoryear{Komossa et al.}{2004}]{kh04}
Komossa, S.; Halpern, J.; Schartel, N.; Hasinger, G.; Santos-Lleo, M.; Predehl, P, 2004, ApJ Letter, 603, 17
\bibitem[\protect\citeauthoryear{Komossa}{2015}]{ko15}
Komossa, S., 2015, JHEAp, 7, 148
\bibitem[\protect\citeauthoryear{Kozlowski et al.}{2010}]{koz10}
Kozlowski, S.; Kochanek, C. S.; Udalski, A.; et al., 2010, ApJ, 708, 927
\bibitem[\protect\citeauthoryear{Kozlowski}{2017}]{ks17}
Kozlowski, S., 2017, A\&A, 597, 128
\bibitem[\protect\citeauthoryear{Lee et al.}{2020}]{lh20}
Lee, C. H.; Hung, T.; Matheson, T.; et al., 2020, ApJL, 892, 1
\bibitem[\protect\citeauthoryear{Li \& Cao}{2008}]{lc08}
Li, S. L.; Cao, X. W., 2008, MNRAS Letter, 387, 41
\bibitem[\protect\citeauthoryear{Leloudas et al.}{2016}]{lf16}
Leloudas, G.; Fraser, M.; Stone, N. C., et al., 2016, Natur Astronomy, 1, 2
\bibitem[\protect\citeauthoryear{Liu et al.}{2017}]{lz17}
Liu, F. K.; Zhou, Z. Q.; Cao, R.; Ho, L. C.; Komossa, S., 2017, MNRAS Letter, 472, 99
\bibitem[\protect\citeauthoryear{Liu et al.}{2020}]{zd20}
Liu, Z.; Li, D.; Liu, H.-Y., et al., 2020, ApJ, 894, 93
\bibitem[\protect\citeauthoryear{Lodato, King \& Pringle}{2009}]{lk09}
Lodato, G.; King A. R.; Pringle J. E., 2009, MNRAS, 392, 332
\bibitem[\protect\citeauthoryear{Loeb \& Ulmer}{1997}]{lu97}
Loeb, A.; Ulmer, A., 1997, ApJ, 489, 573
\bibitem[\protect\citeauthoryear{Lodato et al.}{2015}]{lf15}
Lodato, G.; Franchini, A.; Bonnerot, C.; Rossi E. M., 2015, JHEAp, 7, 158
\bibitem[\protect\citeauthoryear{Lu et al.}{2016}]{ld16}
Lu, K., Du, P., Hu, C., et al., 2016, ApJ, 827, 118
\bibitem[\protect\citeauthoryear{Lynch \& Ogilvie}{2021}]{lo21}
Lynch, E. M.; Ogilvie, G. I., 2021, MNRAS, 501, 5500
\bibitem[\protect\citeauthoryear{MacLeod et al.}{2010}]{mi10}
MacLeod C. L.; Ivezic, Z.; Kochanek, C. S.; et al., 2010, ApJ, 721, 1014
\bibitem[\protect\citeauthoryear{Madejski \& Sikora}{2016}]{ms16}
Madejski, G.; Sikora, M., 2016, ARA\&A, 54, 725
\bibitem[\protect\citeauthoryear{Markwardt}{2009}]{mpf09}
Markwardt, C. B., 2009, Astronomical Data Analysis Software and Systems XVIII ASP Conference Series, Vol. 411, 
proceedings of the conference held 2-5 November 2008 at Hotel Loews Le Concorde, Quebec City, QC, Canada. Edited by David 
A. Bohlender, Daniel Durand, and Patrick Dowler. San Francisco: Astronomical Society of the Pacific, p.251
\bibitem[\protect\citeauthoryear{Merloni et al.}{2015}]{md15}
Merloni, A.; Dwelly, T.; Salvato, A. G.; et al., 2015, MNRAS, 452, 69
\bibitem[\protect\citeauthoryear{Mockler, Guillochon \& Ramirez-Ruiz}{2019}]{mg19}
Mockler, B.; Guillochon, J.; Ramirez-Ruiz, E., 2019, ApJ, 872, 151
\bibitem[\protect\citeauthoryear{Moreno et al.}{2019}]{mv19}
Moreno, J.; Vogeley, M. S.; Richards, G. T.; Yu, W., 2019, PASP, 131, 3001
\bibitem[\protect\citeauthoryear{Mushotzky et al.}{2011}]{me11}
Mushotzky, R. F.; Edelson, R.; Baumgartner, W.; Gandhi, P., 2011, ApJL, 743, 12
\bibitem[\protect\citeauthoryear{Netzer}{2020}]{nh20}
Netzer, H., 2020, MNRAS, 488, 5185
\bibitem[\protect\citeauthoryear{Panagiotou et al.}{2022}]{pp22}
Panagiotou, C.; Papadakis, I.; Kara, E.; Kammoun, E.; Dovciak, M., 2022, ApJ accepted, arXiv:2207.04917
\bibitem[\protect\citeauthoryear{Pancoast et al.}{2014}]{pb14}
Pancoast, A.; Brewer, B. J.; Treu, T., et al., 2014, MNRAS, 445, 3073
\bibitem[\protect\citeauthoryear{Parkinson et al.}{2020}]{pk20}
Parkinson, E. J.; Knigge, C.; Long, K. S., et al., 2020, MNRAS, 494, 4914
\bibitem[\protect\citeauthoryear{Pechacek et al.}{2013}]{pg13}
Pechacek, T.; Goosmann, R. W.; Karas, V.; Czerny, B.; Dovciak M., 2013, A\&A, 556, 77
\bibitem[\protect\citeauthoryear{Peterson et al.}{2002}]{pe02}
Peterson, N. M.; Berlind, P.; Bertram, R.; et al., 2002, ApJ, 581, 197
\bibitem[\protect\citeauthoryear{Peterson et al.}{2004}]{pf04}
Peterson, B. M.; Ferrarese, L.; Gilbert, K. M.; et al., 2004, ApJ, 613, 682
\bibitem[\protect\citeauthoryear{Rees}{1984}]{mr84}
Rees, M. J., 1984, ARA\&A, 22, 471
\bibitem[\protect\citeauthoryear{Rees}{1988}]{re88}
Rees, M. J., 1988, Nature, 333, 523
\bibitem[\protect\citeauthoryear{Richards et al.}{2006}]{rg06}
Richards, G. T.; Lacy, M.; Storrie-Lombardi, L. J.; et al., 2006, ApJS, 166, 470
\bibitem[\protect\citeauthoryear{Rumbaugh et al.}{2018}]{rs18}
Rumbaugh, N.; Shen, Yue; Morganson, E.; et al., 2018, ApJ, 854, 160
\bibitem[\protect\citeauthoryear{Sanchez-Saez et al.}{2018}]{sl18}
Sanchez-Saez, P.; Lira, P.; Mejia-Restrepo, J., et al., 2018, ApJ, 864, 87
\bibitem[\protect\citeauthoryear{Sartori et al.}{2018}]{ss18}
Sartori, Lia F.; Schawinski, Kevin; Trakhtenbrot, B., et al., 2018, MNRAS Letter, 476, 34
\bibitem[\protect\citeauthoryear{Sazonov et al.}{2021}]{sg21}
Sazonov, S.; Gilfanov, M.; Medvedev, P.; et al., 2021, MNRAS, 508, 3820
\bibitem[\protect\citeauthoryear{Sheng, Ross \& Nicholl}{2022}]{sr22}
Sheng, X.; Ross, N.; Nicholl, M., 2022, MNRAS, 512, 5580
\bibitem[\protect\citeauthoryear{Shen et al.}{2011}]{sh11}
Shen, Y.; Richards, G. T.; Strauss, M. A.; et al., 2011, ApJS, 194, 45
\bibitem[\protect\citeauthoryear{Short et al.}{2020}]{sn20}
Short, P.; Nicholl, M.; Lawrence, A.; Gomez, S.; et al., 2020, MNRAS, 498, 4119
\bibitem[\protect\citeauthoryear{Simm et al.}{2016}]{ss16}
Simm, T.; Salvato, M.; Saglia, R.; et al., 2016, A\&A, 585, 129
\bibitem[\protect\citeauthoryear{Stein et al.}{2021}]{sv21}
Stein, R.; van Velzen, S.; Kowalski, M.; et al., 2021, Nature Astronomy, 5, 510
\bibitem[\protect\citeauthoryear{Stone et al.}{2019}]{st19}
Stone, N. C.; Kesden, M.; Chang, R. M.; van Velzen, S.; General Relativity and 
	Gravitation, 2019, 51, 30, arXiv:1801.10180
\bibitem[\protect\citeauthoryear{Stone et al.}{2022}]{ss22}
Stone, Z.; Shen, Y.; Burke, C. J., et al., 2022, MNRAS, 514, 164
\bibitem[\protect\citeauthoryear{Suberlak, Ivezic \& MacLeod}{2021}]{si21}
Suberlak, K. L.; Ivezic, Z.; MacLeod, C., 2021, ApJ, 907, 96
\bibitem[\protect\citeauthoryear{Tachibana et al.}{2020}]{tg20}
Tachibana, Y.; Graham, M. J.; Kawai, N.; Djorgovski, S. G.; Drake, A. J.; Mahabal, A. A.; Stern, D., 2020, ApJ, 903, 54
\bibitem[\protect\citeauthoryear{Takata, Mukuta \& Mizumoto}{2018}]{tm18}
Takata, T.; Mukuta, Y.; Mizumoto, Y., 2018, ApJ, 869, 178
\bibitem[\protect\citeauthoryear{Torricelli-Ciamponi et al.}{2000}]{tf00}
Torricelli-Ciamponi, G.; Foellmi, C.; Courvoisier, T. J. L.; Paltani S., 2000, A\&A, 358, 57
\bibitem[\protect\citeauthoryear{Tout}{1996}]{tp96}
Tout, C. A.; Pols, O.; Eggleton, P.; Han, Z., 1996, MNRAS, 281, 257
\bibitem[\protect\citeauthoryear{Tremaine et al.}{2002}]{tr02}
Tremaine, S., Gebhardt, K., Bender, R., et al. 2002, ApJ, 574, 740
\bibitem[\protect\citeauthoryear{Ulrich,  Maraschi \& Urry}{1997}]{um97}
Ulrich, M. H.; Maraschi, L.; Urry C. M., 1997, ARA\&A, 35, 445
\bibitem[\protect\citeauthoryear{van Velzen et al.}{2011}]{ve11}
van Velzen, S.; Farrar, G. R.; Gezari, S.; et al., 2011, ApJ, 741, 73
\bibitem[\protect\citeauthoryear{van Velzen et al.}{2019}]{vv19}
van Velzen, S.; Gezari S.; Cenko, S. B.; et al., 2019, ApJ, 872, 198
\bibitem[\protect\citeauthoryear{van Velzen et al.}{2021}]{vg21}
van Velzen, S.; Gezari, S.; Hammerstein, E.; et al., 2021, ApJ, 908, 4
\bibitem[\protect\citeauthoryear{Wagner \& Witzel}{1995}]{ww95}
Wagner, S. J.; Witzel, A., 1995, ARA\&A, 33, 163
\bibitem[\protect\citeauthoryear{Wang et al.}{2018}]{wy18}
Wang, T.; Yan, L.; Dou, L., et al., 2018, MNRAS, 477, 2943
\bibitem[\protect\citeauthoryear{Williams et al.}{2020}]{wp20}
Williams, P. R.; Pancoast, A.; Treu, T.; et al., 2020, ApJ, 902, 74
\bibitem[\protect\citeauthoryear{Wyrzykowski et al.}{2017}]{wz17}
Wyrzykowski, L.; Zielinski, M.; Kostrzewa-Rutkowska, Z.; et al., 2017, MNRAS Letter, 465, L114
\bibitem[\protect\citeauthoryear{Yan \& Xie}{2018}]{yx18}
Yan, Z.; Xie, F., 2018, ApJ, 475, 1190
\bibitem[\protect\citeauthoryear{Zanazzi \& Ogilvie}{2020}]{zo20}
Zanazzi, J. J.; Ogilvie, G. I., 2020, MNRAS, 499, 5562
\bibitem[\protect\citeauthoryear{Zhang et al.}{2022}]{zs22}
Zhang, W. J.; Shu, X. W.; Sheng, Z. F., et al., 2022, A\&A, 660, 119
\bibitem[\protect\citeauthoryear{Zhang \& Feng}{2017}]{zh17}
Zhang, X. G.; Feng L. L., 2017, MNRAS, 464, 2203
\bibitem[\protect\citeauthoryear{Zhang et al.}{2021a}]{zh21a}
Zhang, X. G.; Zhang, Y., Cheng, P., et al., 2021a, ApJ, 922, 248
\bibitem[\protect\citeauthoryear{Zhang}{2021b}]{zh21b}
Zhang, X. G., 2021b, MNRAS Letter, 500, 57
\bibitem[\protect\citeauthoryear{Zhang}{2022}]{zh22}
Zhang, X. G., 2022, ApJ, 928, 182
\bibitem[\protect\citeauthoryear{Zhang}{2022b}]{zh22b}
Zhang, X. G., 2022b, MNRAS Letter, 516, 66, Arxiv:2208.05253
\bibitem[\protect\citeauthoryear{Zhang}{2022c}]{zh22c}
Zhang, X. G., 2022c, MNRAS Letter, 517, 71, Arxiv:2209.09037
\bibitem[\protect\citeauthoryear{Zhang}{2023}]{zh23}
Zhang, X. G., 2023, ApJ accepted, Arxiv:2309.00852
\bibitem[\protect\citeauthoryear{Zhou et al.}{2021}]{zl21}
Zhou, Z. Q.; Liu, F. K.; Komossa, S., et al., 2021, ApJ, 907, 77
\bibitem[\protect\citeauthoryear{Zu, Kochanek \& Peterson}{2011}]{zu11}
Zu, Y.; Kochanek, C. S.; Peterson, B. M., 2011, ApJ, 735, 80
\bibitem[\protect\citeauthoryear{Zu et al.}{2013}]{zu13}
Zu, Y.; Kochanek, C. S.; Kozlowski, S.; Udalski, A., 2013, ApJ, 765, 106
\end{thebibliography}
\end{document}